\documentclass{article}
	\usepackage{arxiv}
	
	\usepackage[utf8]{inputenc} % allow utf-8 input
	\usepackage[T1]{fontenc}    % use 8-bit T1 fonts
	\usepackage{hyperref}       % hyperlinks
	\usepackage{url}            % simple URL typesetting
	\usepackage{booktabs}       % professional-quality tables
	\usepackage{amsfonts}       % blackboard math symbols
	\usepackage{nicefrac}       % compact symbols for 1/2, etc.
	\usepackage{microtype}      % microtypography
	\usepackage{graphicx}
	
	\usepackage{mathrsfs}
	\hypersetup{hidelinks}
	\usepackage[hyphenbreaks]{breakurl}
	\usepackage{multirow}
	\usepackage{subfig}
	\usepackage{lipsum}
	\usepackage{array}
	\usepackage{enumitem}
	\usepackage{amsmath,amsfonts,amssymb,amsthm}
	\usepackage[ruled,linesnumbered]{algorithm2e}
	
	\theoremstyle{definition}
	\newtheorem{defn}{Definition}[section]
	\allowdisplaybreaks
	\DeclareMathAlphabet{\mathcal}{OMS}{cmsy}{m}{n}
	
	\title{ReAD: A Regional Anomaly Detection Framework Based on Dynamic Partition}
	\author{Huaishao Luo$^{1}$, Chuishi Meng$^{2}$, Bowen Wu$^{2}$, Junbo Zhang$^{2}$, Tianrui Li$^{1}$, Yu Zheng$^{2}$\\
		$^1$Southwest Jiaotong University, Chengdu, China \\
		{\tt huaishaoluo@gmail.com, trli@swjtu.edu.cn} \\
		$^2$Urban Computing Business Unit, JD Digits, Beijing, China \\
		{\tt meng.chuishi@jd.com, wbw@mail.nankai.edu.cn, } \\
		{\tt jbzhang86@gmail.com, msyuzheng@outlook.com }
  	}

\begin{document}
		\maketitle
		\begin{abstract}
			The detection of the abnormal area from urban data is a significant research problem. However, to the best of our knowledge, previous methods designed on spatio-temporal anomalies are road-based or grid-based, which usually causes the data sparsity problem and affects the detection results. In this paper, we proposed a dynamic region partition method to address the above issues. Besides, we proposed an unsupervised REgional Anomaly Detection framework (ReAD) to detect abnormal regions with arbitrary shapes by jointly considering spatial and temporal properties. Specifically, the proposed framework first generate regions via a dynamic region partition method. It keeps that observations in the same region have adjacent locations and similar non-spatial attribute readings, and could alleviate data sparsity and heterogeneity compared with the grid-based approach. Then, an anomaly metric will be calculated for each region by a regional divergence calculation method. The abnormal regions could be finally detected by a weighted approach or a wavy approach according to the different scenario. Experiments on both the simulated dataset and real-world applications demonstrate the effectiveness and practicability of the proposed framework.
		\end{abstract}

		\keywords{Dynamic region partition \and Regional anomaly detection \and Spatio-temporal anomaly detection \and Urban computing}
		
		% ================>Introduction<===================
		\section{Introduction}
		\label{sec_introduction}
		Anomaly detection is widely used in many scenarios, e.g., credit card fraud and industrial damage detection \cite{Chandola2009}. Most of them are usually associated with scalar datasets or time series. As a branch of anomaly detection, urban anomaly detection, which also belongs to urban computing \cite{Zheng2014}, mainly focuses the anomalies in the urban. In urban computing, observations generally contain two kinds of descriptive information: spatial attribute, namely geographic coordinates, and temporal attribute, which shows as the dynamic readings with time. Formatively, each observation is a triplet $\langle$geographic locations, timestamp, readings$\rangle$. Thus, urban anomaly detection should consider spatial and temporal attributes synthetically. It is important to public safety and urban policy-making. For instance, if a crowd gathering can be detected timely, the risks to public safety may be decreased beforehand. One negative example mentioned in \cite{Zhang2018} is that a tragic stampede resulting in 36 people killed and 49 injured took place in Shanghai. There are more than 300,000 people flocked for a popular light show on New Year’s Eve (Dec. 31th, 2014). The declaration from the authority reflects that the correct estimation of crowd size and well-preparation can avoid this event.

		Existing urban anomaly detection methods can be divided into the point-based method, line-based method, and region-based method by their detection objective. The point-based method is usually used to detect abnormal points. As an instance, spatial outliers detection is a typical scenario \cite{Shi2016,Zheng2017}. Likewise, the trajectory anomalies detection can regard as a type of line-based method \cite{Ge2010,Liu2011,Li2006,Li2007}. The target of the region-based method is to find group anomalies or collective anomalies \cite{Zheng2015,Zhang2018}. Our goal is to detect regional anomalies, which relates to the region-based method.
		
\begin{figure}[tbp]
			\centering
			\subfloat[Road based partition] {
				\includegraphics[width=1.6in]{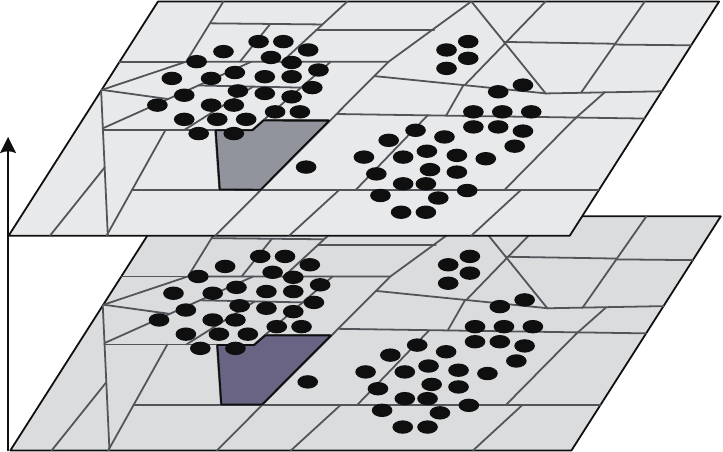}
				\label{fig_roadBasedPartition}
			}
			\hspace{.1in}
			\subfloat[Grid based partition] {
				\includegraphics[width=1.6in]{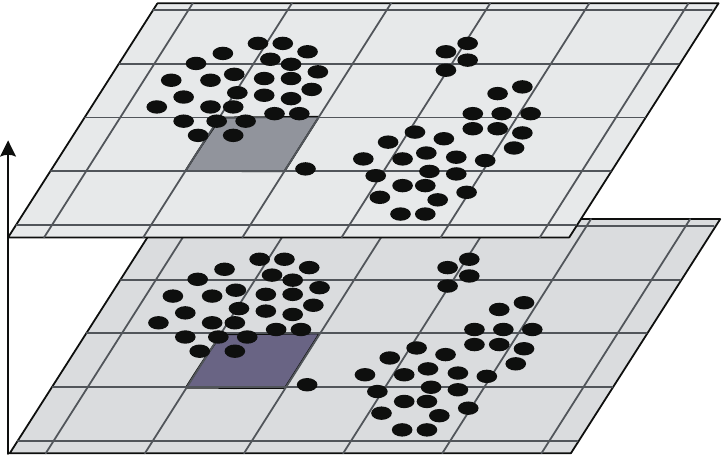}
				\label{fig_gridBasedPartition}
			}
			\hspace{.1in}
			\subfloat[Dynamic partition] {
				\includegraphics[width=1.6in]{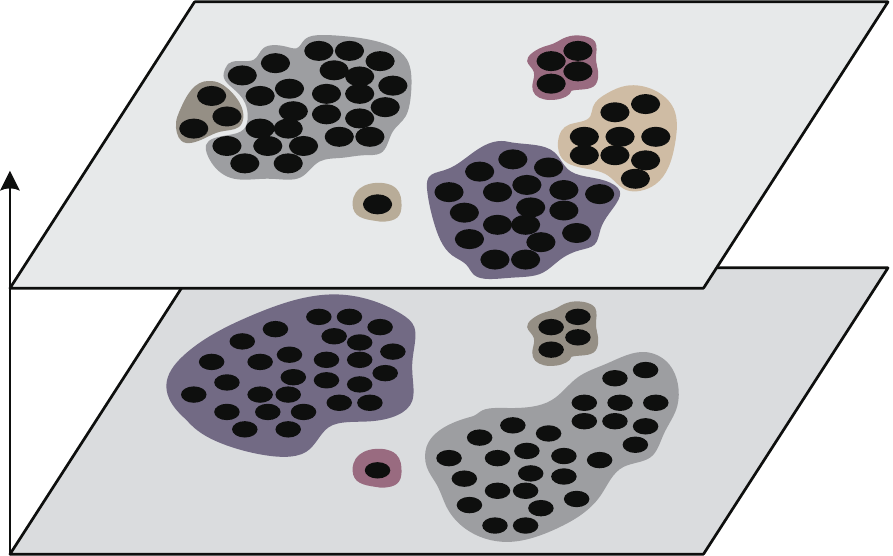}
				\label{fig_dynamicPartition}
			}
			\caption{Different type of spatial partition.}
			\label{fig_partition_type}
		\end{figure}

		However, the previous region-based detection method usually involves fixed-based regional partition. Both \cite{Zheng2015} and \cite{Zhang2018} partition a city into some regions by major roads, e.g., highways and arterial roads. Such a partition approach we refer to as a road-based method shown in Figure \ref{fig_roadBasedPartition}. Another fixed-based regional partition is grid-based partition shown as Figure \ref{fig_gridBasedPartition}, which usually used in some applications in the urban data mining, e.g., the crowd flows prediction \cite{Zhang2017}. The grid-based method partitions a city into a grid map based on the longitude and latitude where a grid denotes a region. The big problem caused by the fixed-based regional partition is data sparsity because the distribution of observations is unbalanced in the whole area, which will influence the detection results. Another issue is that the heterogeneity of data will also affect detection \cite{Shi2017}.

		To alleviate the influence of data sparsity and heterogeneity, we propose a novel method to partition the area dynamically. Figure \ref{fig_dynamicPartition} illustrates the process of the dynamic partition, in which the area number and shape are changed with the time flow. The dynamic partition operates on locations and readings directly. Thus, the data sparsity issue is negligible compared with that caused by fixed-based partition method. The generated regions need to satisfy two principles: locations are adjacent, and readings are similar. Thus, data heterogeneity is not a highlighted problem anymore.

		The target of regional anomalies detection is to detect whether a region is significantly different from others. Besides exploring the regional partition approach, another critical question is to exploring a metric to measure the abnormality. In urban data, this metric is supposed to have a relationship with the temporal attributes or both spatial and temporal attributes. On the one hand, we can not assume an observation is abnormal by only judging its value is lower or higher than others in spatial distribution. E.g., the bike drivers in the center of a city are always much more than the ambient areas. Thus it is unreasonable to regard an area with a lower or higher driver number as an abnormal area. On the other hand, using only the fluctuation of the time series to judge whether a timestamp is abnormal is also not enough. The reason is there are too many factors like workday or raining weather that will influence the readings. Inspired by the study of \cite{Barz2018}, we adopt the divergence as the metric. It is calculated considering the readings belonging to both a target region and its surrounding regions. The difference with \cite{Barz2018} is the divergence plays two roles in the paper. Besides the direct anomaly metric, the divergence is also used as a weight. So, we also call the divergence as relative divergence.
		
		The choice that using the relative divergence to describe the fundamental change caused by an unusual event is adaptive. As an instance, although the number of bike drivers in a city is different on the raining day and sunny day, each station should have a stable relative divergence under these weather conditions. A constant relative divergence means each station has the same trend: riders are less on a raining day and more on a sunny day as common sense. If the drivers of a station have no such tendency, we believe there is an anomaly. E.g., a concert that causes a big riders' number. In other words, when the relative divergence changes massively, there may be an anomaly we want to track. 

		Combining dynamic regions and the relative divergence, we propose an unsupervised REgional Anomaly Detection framework, entitled ReAD, to detect regional anomalies in this paper. Explicitly, we first partition the observations using its geographic locations and readings. Each of the generated regions should always hold the following two principles: locations are adjacent, and readings are similar. Then, we calculate the relative divergence of each region. Finally, finding the anomalies by the calculated divergence. These processes keep the final anomalies are regional and deviate from the normal regionals with comparable values.
		
		Our proposed framework involves dynamic region partition instead of the fixed-based partition for the following advantages:
		\begin{itemize}[leftmargin=2em]
			\item[1)] Dynamic region partition gathers similar readings within the nearby location together directly. It is capable of overcoming the data-sparse in the fixed-based partition method and addressing the heterogeneity of the given data.
			\item[2)] Dynamic region partition has no grid size choice to affect the results of detection. However, the grid size is essential in the grid-based partition. It also no need to combine the final nearby detected regions, which should be operated in the road-based approach.
		\end{itemize}

		Temporal attributes play two roles in the proposed framework for different applications. One is to assist the detection considering some scenarios emphasize spatial anomalies, e.g., credit anomaly detection. Another is to investigate the fluctuation of the divergence. The reason is most situations can not generate anomalies without temporal information, e.g., bike drivers anomaly detection. Thus, we design two approaches to generate the final regional anomalies. Correspondingly, one is called weighted approach, and another is called wavy approach.

		Our contributions are three-fold:
		\begin{itemize}[leftmargin=2em]
			\item A partition method is proposed to partition regions dynamically. Each generated region satisfies two principles: locations are adjacent, and readings are similar.
			\item An unsupervised framework ReAD is proposed to address regional anomaly detection. It heuristically involves dynamic region partition, regional divergence calculation, and anomalous region generation.
			\item We evaluate the proposed framework by a synthetic spatio-temporal dataset and two real-world applications. The results demonstrate the effectiveness and practicability of the framework.
		\end{itemize}

		In the following, we first present the preliminary notations and definitions, and the overview of the proposed framework ReAD in Section \ref{sec_overview}. Then, we elaborate on the proposed framework in Section \ref{sec_methodology}. The experiments and analysis are discussed in Section \ref{sec_experiments}, followed by an introduction to the related work in Section \ref{sec_related_work}. Finally, we conclude the paper in Section \ref{sec_conclusion}.

		% ================>Model Description<=====================
		\section{Overview}
		\label{sec_overview}
		In this section, we introduce some preliminary notations, definitions, and the main structure of the proposed framework.
		
		\subsection{Preliminary}
		\label{sec_preliminary}
		\begin{defn}[Region]
			In this study, all observations $s_t$ at the time slot $t$ are partitioned into some regions $r_t=\left\{r_{t,1}, r_{t,2}, \cdots, r_{t,n_t} \right\}$, where $n_t$ is the number of regions at time slot $t$. Each region needs to satisfy the following two criteria:
			\begin{itemize}[leftmargin=*]
				\item[1)] Adjacent locations,
				\item[2)] Similar readings: $\forall v_i, v_j \in r_{t,k}$, $dist(v_i, v_j) \leq \delta_d$.
			\end{itemize}
		\end{defn}
		\begin{defn}[Regional Anomaly]
			A regional anomaly is a region that deviates significantly from the nearby regions or whole regions. In other words, the distribution of each point in a region deviates significantly from other points outside it.
		\end{defn}
		\noindent
		\textbf{Problem Definition}. Given a data stream $s$, each element of it is a triplet $\langle$ $l$, $t$, $v$ $\rangle$ representing that a value $v$ is observed in geographic coordinate (i.e., longitude, latitude) $l$ at time slot $t$. We aim to detect $\varrho$ anomalous regions $\mathcal{A}_t=\left\{r_{t,1}, r_{t,2}, \cdots, r_{t,\varrho} \right\}$ at time epoch $t$.
		\begin{figure}[htbp]
			\centering
			\subfloat[] {
				\includegraphics[width=1.5in]{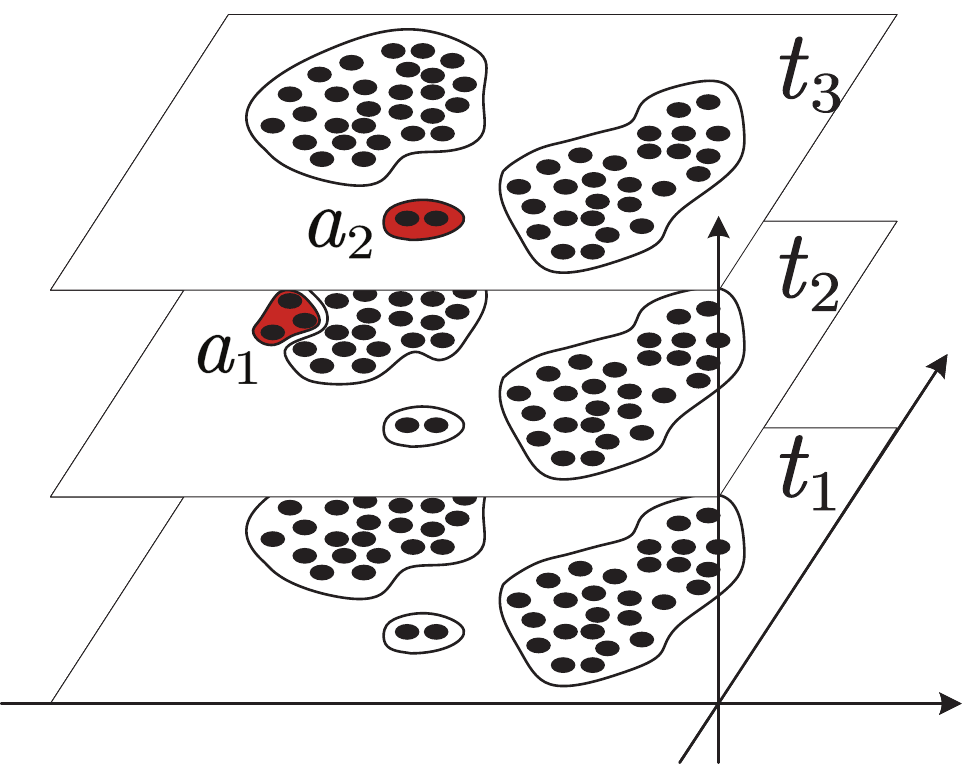}
				\label{fig_problemA}
			}
			\hspace{.1in}
			\subfloat[] {
				\includegraphics[width=4.1in]{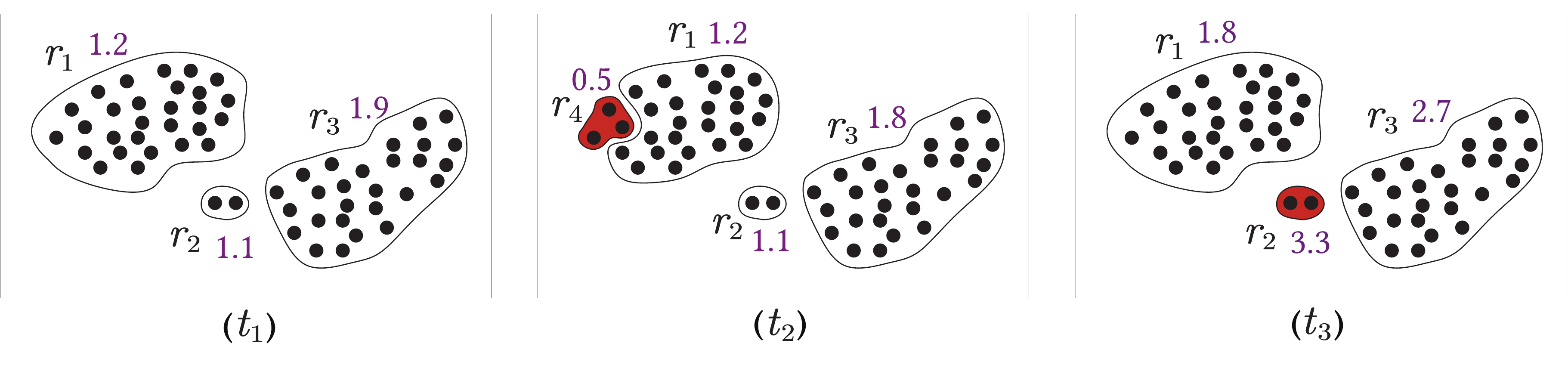}
				\label{fig_problemB}
			}
			\caption{An instance of the problem definition, the numbers colored with purple are the mean value of each region (Best seen in color).}
			\label{fig_problem}
		\end{figure}
		
		Figure \ref{fig_problem} is an instance to illustrate the problem definition. There are two regional anomalies ($a_i, i=1,2$) detected from three consecutive time interval [$t_1, t_3$] in Figure \ref{fig_problemA}. Figure \ref{fig_problemB} elaborates the locations and readings of each timestamp. The numbers colored with purple are the mean value of each region. As it shows, the regions are partitioned dynamically by the geo-coordinates and readings of observations at each time slot. E.g., all observations are partitioned into three regions at time slot $t_1$. However, there are four regions at time slot $t_2$, because the readings in $r_4$ have changed. The proposed method detects abnormal regions combining spatial and temporal attributes. For example, $r_4$ at time slot $t_2$ is an unusual region since their readings entirely different from the readings of region $r_1$. Besides, $r_2$ at $t_3$ is an abnormal region because the readings go up sharply compared with the other two regions.
		
		As a conclusion, we can not identify the anomalies just by spatial attribute and also can not identify the anomalies only by temporal attribute. The task of regional anomaly detection involves both of them.
		\begin{figure}[thbp]
			\centering
			\includegraphics[width=.98\textwidth]{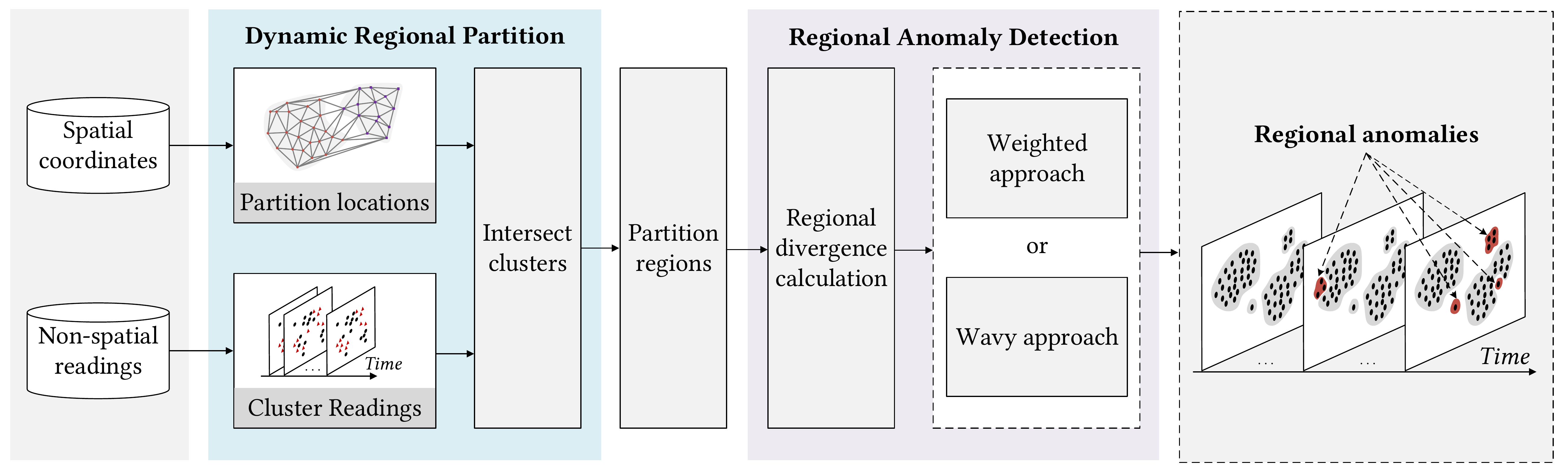}
			\caption{The overall structure of our proposed framework.}
			\label{fig_framework}
		\end{figure}

		\subsection{Framework}
		\label{sec_framework}
		The main structure of our proposed framework is illustrated in Figure \ref{fig_framework}. The two main parts are dynamic region partition and regional anomaly detection. For the dynamic region partition, the key idea is to partition locations and cluster readings at each time slot, respectively. Then, an intersection operation is carried out to obtain final regions. These two steps keep locations adjacent and readings similar in each generated region. To the best of our knowledge, it is the first time to use intersection operation in the regional partition. It is a concise and efficient approach to get the target regions. For regional anomaly detection, we first calculate a divergence for each region. Then, two approaches are proposed to address different scenarios. A weighted method is used in spatial anomaly regions detection, in which the weights are calculated considering temporal information. A wavy approach is proposed to spatio-temporal anomaly detection. All these steps will be described in detail in Section \ref{sec_methodology}.

		\section{Methodology}
		\label{sec_methodology}
		In this section, we elaborate regional partition, regional divergence calculation, and anomalous region generation of the proposed framework shown as Figure \ref{fig_framework}.
		
		\subsection{Regional Partition}
		\label{sec_regional_partition}
		Regional partition is an essential part of the framework. Its output is regions, in which observations have adjacent locations and similar readings. It is challenging to consider locations and readings simultaneously. The reason is the location of an observation is always fixed, but its reading will change with time. The location and reading have different features. From a clustering perspective, the task of regional partition essentially is a cluster problem, which can be addressed by unsupervised cluster algorithms. Putting above two knots together, we develop an effective method to solve the problem of regional partition. The technique first cluster them individually, then intersect these clusters. Thus, each of the final regions satisfies the principles of adjacent locations and similar readings.
		
		Before introducing the partition approach, it needs to explain the reason why we need a novel partition method. The critical point is that the partition regions we need are arbitrarily shaped and should form a full coverage for all observations. It makes most of the cluster methods are infeasible. As an instance, the density-based clusters (e.g., DBSCAN \cite{Ester1996}) can cluster arbitrarily shaped regions but can not ensure them cover all observations. It is also a challenge to choose proper parameters (e.g., the $MinPts$ in DBSCAN). The expectation of arbitrary shape and full coverage makes a customized cluster approach necessary.
		
		 \textbf{Partition locations}. Inspired by CFDP (Clustering by Fast Search and Find of Density Peaks) \cite{Rodriguez2014}, we propose a partition algorithm (or cluster algorithm) based on Delaunay triangulation \cite{Tsai1993} to address above challenges. 
		
		 A Delaunay triangulation is also known as a Delone triangulation. Its objective is to maximize the minimum angle of all the angles of the triangles in the triangulation. We use Delaunay triangulation mainly because it is parameters free.
		\begin{figure}[htbp]
			\centering
			\subfloat[Delaunay triangulation] {
				\centering
				\includegraphics[width=0.237\textwidth]{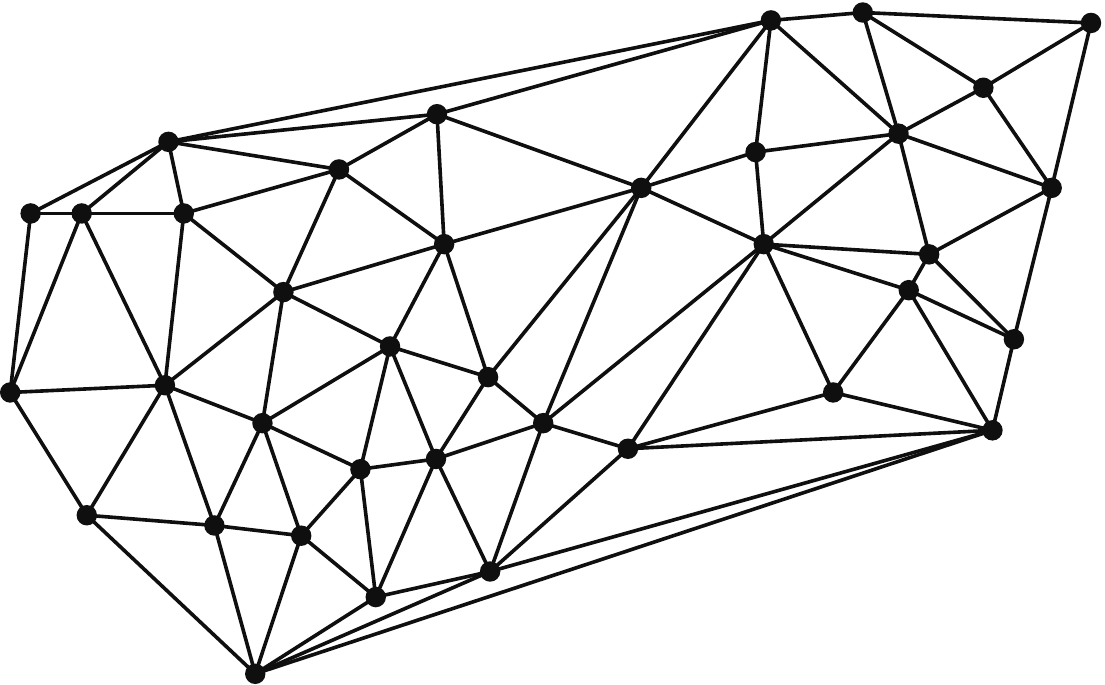}
				\label{fig_Delaunay_triangulationA}
			}
			\subfloat[Density calculation] {
				\centering
				\includegraphics[width=0.24\textwidth]{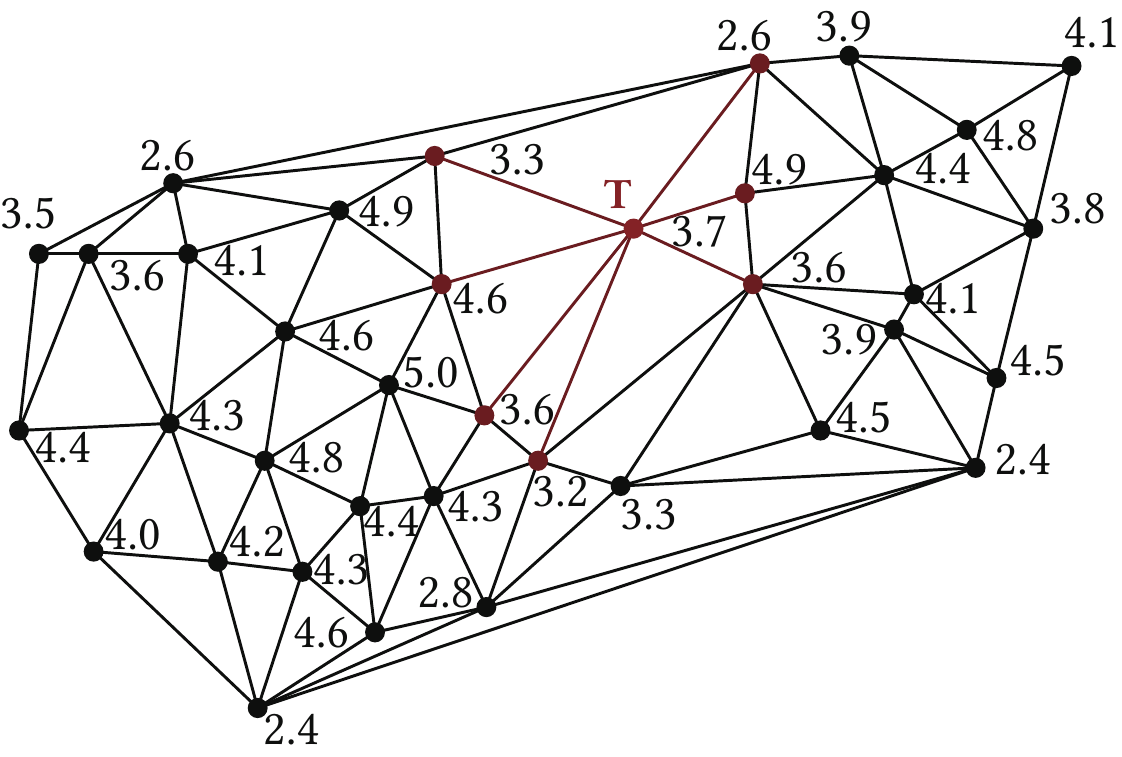}
				\label{fig_Delaunay_triangulationB}
			}
			\subfloat[Partition process] {
				\centering
				\includegraphics[width=0.24\textwidth]{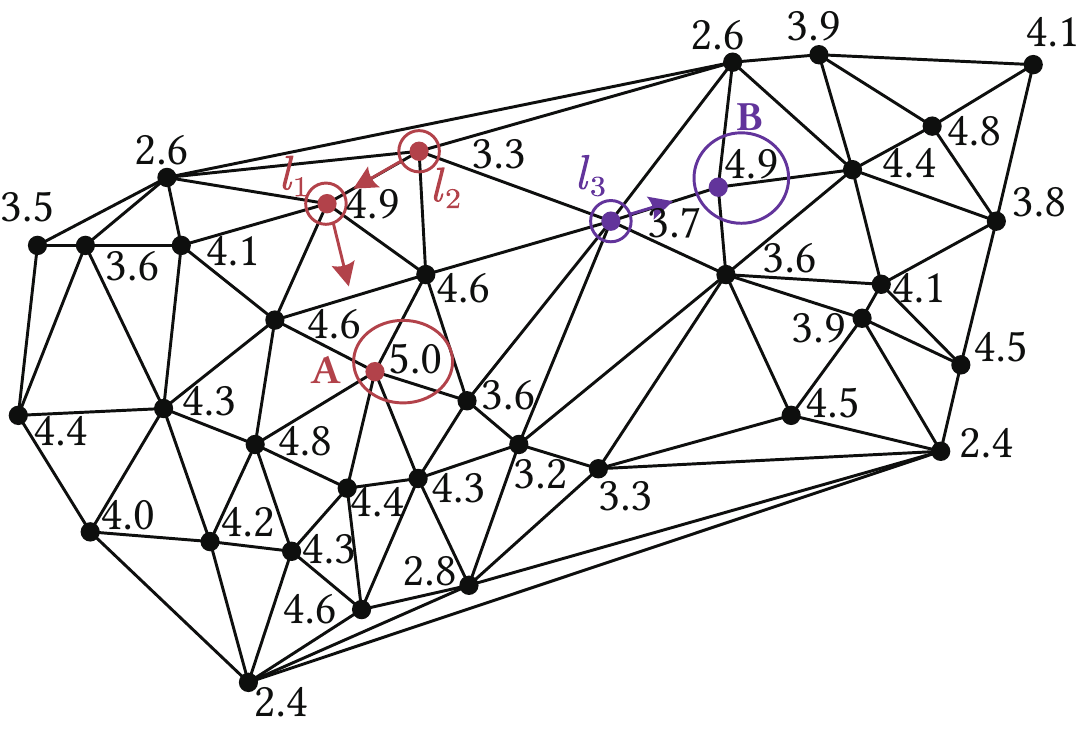}
				\label{fig_Delaunay_triangulationC}
			}
			\subfloat[Region partition results] {
				\centering
				\includegraphics[width=0.24\textwidth]{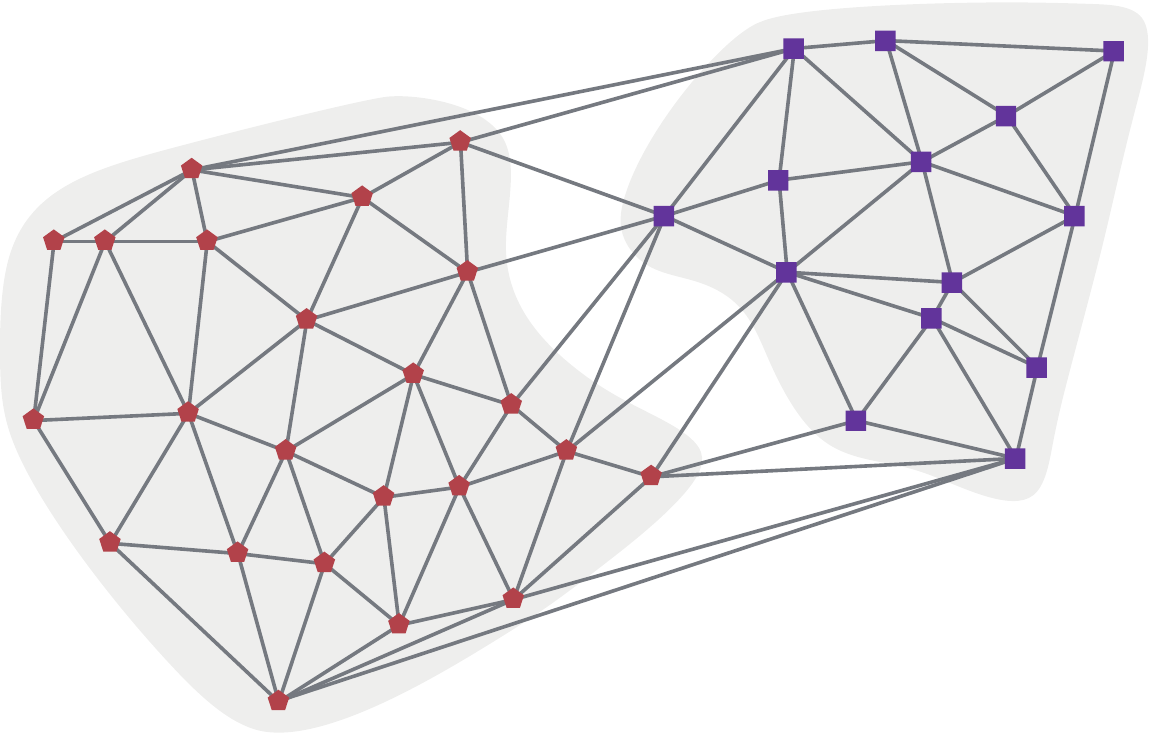}
				\label{fig_Delaunay_triangulationD}
			}
			\caption{Partition algorithm on locations.}
			\label{fig_Delaunay_triangulation}
		\end{figure}

		Figure \ref{fig_Delaunay_triangulation} with four subfigures is an illustration of the proposed partition algorithm. Given some geographic coordinates $s_t$, the generated Delaunay triangulation $\mathcal{T}_{s_t}$ is a connected graph, in which a vertex represents a location, and an edge connects two adjacent locations. Figure \ref{fig_Delaunay_triangulationA} is a diagram of Delaunay triangulation. If we cut the longest edge in sequence, there will be two clusters shown as Figure \ref{fig_Delaunay_triangulationD}. It also follows the rule that the edge length between clusters is longer than the edge length inside. Through analyzing the two clusters, there is a truth that the points in the boundary of a cluster usually have longer edges to connect with other clusters. One step closer, these boundary points typically have a higher variance on the length of edge around themselves than that of other points. Thus, we define the density of each location $l_i$ by:
		\begin{equation}\label{eq_rho}
			\begin{aligned}
				\rho_{l_i} = \log\left(\frac{1}{\sigma_{l_i}}\right),
			\end{aligned}
		\end{equation}
		Let $\eta_i=\left\{l_j, l_{j+1}, \cdots, l_{j+k-1}\right\}$ denote $k$ neighbors of $l_i$ (e.g., the T location in Figure \ref{fig_Delaunay_triangulationB} owns $k=7$ neighbors), the $\sigma_{l_i}$ is calculated with the following equations:
		\begin{align}
			\sigma_{l_i} &= \sqrt{\frac{1}{k} \sum_{{l_j} \in \eta_{l_i}}{\left( e_{l_i,l_j}-\mu_{l_i} \right)^2}}, \\
			\mu_{l_i} &= \frac{1}{k} \sum_{{l_j} \in \eta_{l_i}}{e_{l_i,l_j}}, \\
			e_{l_i,l_j} &= \lVert l_i - l_j \rVert_2, \label{eq_e}
		\end{align}
		where, $\lVert \cdot \rVert_2$ is the Euclidean norm. The numbers (round to two decimal places) in Figure \ref{fig_Delaunay_triangulationB} are calculated densities for each location. As we expect, the density of the points in the boundary is usually smaller than that in the central position. It is note that here using the variance $\sigma_{l_i}$ instead of the average $\mu_{l_i}$ to calculate density $\rho_{l_i}$. The reason is the variance $\sigma_{l_i}$ is stable no matter for sparse or dense points. This feature is useful to alleviate the influence of the data heterogeneity.

		After obtaining the density, the task is to allocate each of the locations to a latent cluster. As the previous conclusion, the density in the central position is usually higher than that in the boundary. So there are three steps to get clusters. Firstly, let each point go with the direction of the density higher than itself. Then, choose the locations with the top density as the center of each cluster. Finally, gather the points to each cluster using the opposite direction with the first step recursively. Following the CFDP, $\delta_{l_i}$ is defined as the minimum distance between the location $l_i$ and any other locations with higher density:
		\begin{equation}\label{eq_delta}
			\delta_{l_{(i)}} = \left\{
			\begin{array}{lr}
				\min\limits_{j<i} \left( e_{l_{(i)}, l_{(j)}} \right), \quad &i \geq 2; \\
				\max\limits_{j \geq 2} \left( \delta_{l_{(j)}} \right), \quad &i = 1.
			\end{array}
			\right.
		\end{equation}
		where, the $l_{(i)}$ is the reordering of $l_i$: $\rho_{l_{(1)}} \geq \rho_{l_{(2)}} \geq \cdots \geq \rho_{l_{(i)}} \geq \cdots$. According to $\delta_{l_i}$, each location owns a parent that has the minimum distance with itself, except the one with the biggest minimum distance.
		
		Figure \ref{fig_Delaunay_triangulationC} illustrates the process of clustering using the density $\rho_{l_i}$ and  the minimum distance $\delta_{l_i}$ of each location. The locations with the biggest $c=2$ readings $\gamma_{l_i}=\rho_{l_i} \times \delta_{l_i}$ are first chosen as the centers of clusters, and then other locations recursion to one of the centers following their parent locations. For example, $A$ and $B$ are the chosen two centers. $l_2$ belongs to $A$ by the path $l_2 \rightarrow l_1 \rightarrow A$. Likewise, $l_3$ belongs to $B$. Figure \ref{fig_Delaunay_triangulationD} illustrates the final clustering results. Benefit from the elaborate density function, the proposed algorithm has fewer parameters than CFDP.

		\textbf{Cluster readings}. Compared with partition locations, cluster readings is a relatively straightforward process. Any cluster method can finish this step because the condition of similar readings is always satisfied in the same cluster. As an alternative approach, we use the CFDP with the default setting to cluster the readings at time slot $t$. Nevertheless, KNN or other cluster methods are also feasible.

\begin{figure}[thbp]
			\centering
			\includegraphics[width=.58\textwidth]{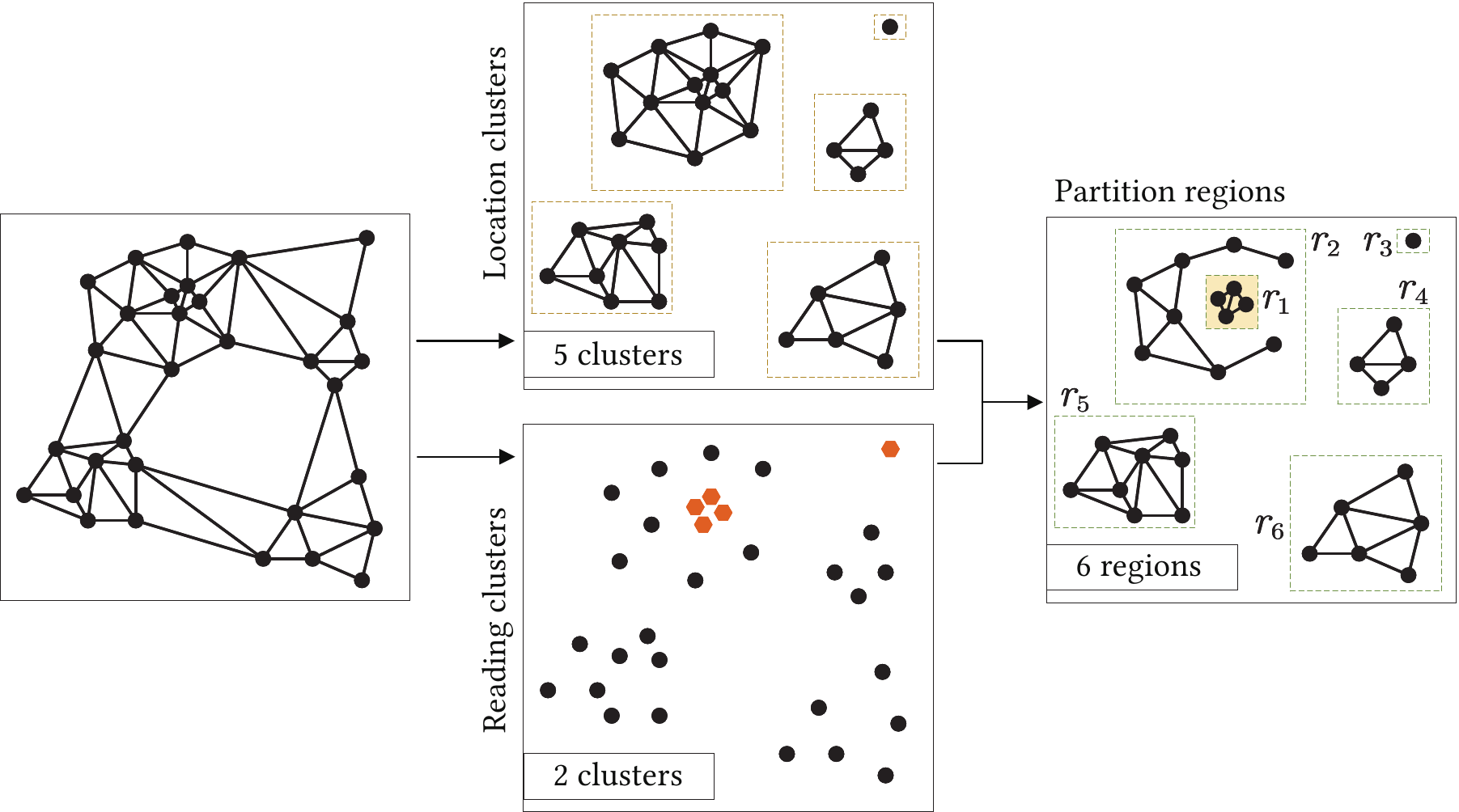}
			\caption{The process of partition regions.}
			\label{fig_partitionProcess} 
		\end{figure}

		\textbf{Intersect clusters}. We denote the location clusters at time slot $t$ by $\mathcal{L} = s_t/l = \left\{c_{t,1}^{(l)}, c_{t,2}^{(l)}, \cdots, c_{t,n_t(l)}^{(l)}\right\}$ and the reading clusters at time slot $t$ by $\mathcal{V}= s_t/v = \left\{c_{t,1}^{(v)}, c_{t,2}^{(v)}, \cdots, c_{t,n_t(v)}^{(v)}\right\}$ where $n_t(l)$ is the location cluster number and $n_t(v)$ is the reading cluster number. Thus, the final partition regions $r_t$ are their intersection:
		\begin{align}
			r_t = \mathcal{L} \cap \mathcal{V} = \left\{r_{t,1}, r_{t,2}, \cdots, r_{t,n_t} \right\},
		\end{align}
		The intersection operation keeps each of the final regions satisfies the principles of adjacent locations and similar readings. Figure \ref{fig_partitionProcess} illustrates the whole partition process. There are 5 location clusters, 2 value clusters, and the 6 final regions in this example.

		The final partition regions are dynamic with different locations or readings. Actually, the location partition almost changeless until the points change their positions (such as increasing the bike stations). At this moment, this partition can be regarded as repartition the points inside some fixed regions according to their readings at each time slot.

		\subsection{Regional Divergence Calculation}
		\label{sec_regional_divergence_calculation}
		We calculate the metric of abnormality by the relative divergence of each region. According to our analysis mentioned in Section \ref{sec_introduction}, the relative divergence is essential when considering the regional anomaly detection. The reason is the external factors (e.g., sunny or raining) except for the unusual event (e.g., a concert) is changing with time. We assume that the relative divergence is stable under the condition of external factors, but drastic fluctuation under the abnormal events. Motivated by the works on collective anomaly detection \cite{Barz2018,Liu2013,Jiang2015}, we use the Kullback-Leiber (KL) divergence to calculate the relative divergence:
		\begin{align}
			\mathfrak{D}_{KL}(p, q) = \mathbb{E}_p\left[-\log \frac{p}{q} \right],
		\end{align}
		Specifically, we measure each region $r_{t,i}$ with the relative divergence from two aspects, one is local divergence, and another is global divergence. The local divergence involves the probability density $p_{r_{t,i}}$ and the distribution $p_{c_{t,i}^{(l)}}$, where $c_{t,i}^{(l)}$ is the location cluster containing the region $r_{t,i}$. The global divergence involves the probability density $p_{r_{t,i}}$ and the distribution $p_{\lnot r_{t,i}}$ of the rest regions except $r_{t,i}$ at time slot $t$. We investigate Kernel Density Estimation (KDE) for calculating these distributions:
		\begin{align}
			p_{\Re}\left( v_\iota \right) = \frac{1}{\left| \Re \right|} \sum\limits_{j \in \Re}{\kappa \left( v_\iota, v_j \right)}, \quad \Re \in \left\{ r_{t,i}, c_{t,i}^{(l)}, \lnot r_{t,i} \right\},
		\end{align}
		where, the kernel is Gaussian kernel:
		\begin{align}
			\kappa \left(v_\iota, v_j\right) = (2\pi\sigma)^{-\frac{1}{2}} \exp(-\frac{\lVert v_\iota - v_j \rVert^2}{2\sigma^2}),
		\end{align}
		Thus, we can generate the regional divergence $\mathfrak{D}_{r_{t,i}}$ with:
		\begin{align}\label{eq_calculate_divergence}
			\mathfrak{D}_{r_{t,i}} = \lambda \cdot \mathfrak{D}_{KL}^{local}(p_{r_{t,i}}, p_{c_{t,i}^{(l)}}) + (1-\lambda) \cdot \mathfrak{D}_{KL}^{global}(p_{r_{t,i}}, p_{\lnot r_{t,i}}),
		\end{align}
		where, $\mathfrak{D}_{KL}^{local}(p_{r_{t,i}}, p_{c_{t,i}^{(l)}})$ is the local divergence (e.g, $\mathfrak{D}_{KL}(p_{r_1}, p_{r_2})$ in Figure \ref{fig_partitionProcess}), and $\mathfrak{D}_{KL}^{global}(p_{r_{t,i}}, p_{\lnot r_{t,i}})$ is the global divergence (e.g., $\mathfrak{D}_{KL}(p_{r_1}, p_{\cup r_{i}, i=2,3,4,5,6})$ in Figure \ref{fig_partitionProcess}). $\lambda \in [0, 1]$ is a trade off between the local and the global divergence. 

		\subsection{Regional Anomalies Detection}
		\label{sec_regional_anomalies_detection}
		The temporal dimension plays a vital role in our framework. We explore two approaches to cover two types of situations. The first one is to use temporal information to help the detection of spatial anomalies where spatial anomalies are conspicuous, e.g., credit anomaly detection. We call this approach as a weighted approach. The second one is to use the fluctuation of the divergence series to decide whether a region is abnormal or not, e.g., NYC bike anomaly detection. In this situation, spatial attributes are not enough to make a decision. We call this approach a wavy approach.
		\begin{figure}[thbp]
			\centering
			\includegraphics[width=.78\textwidth]{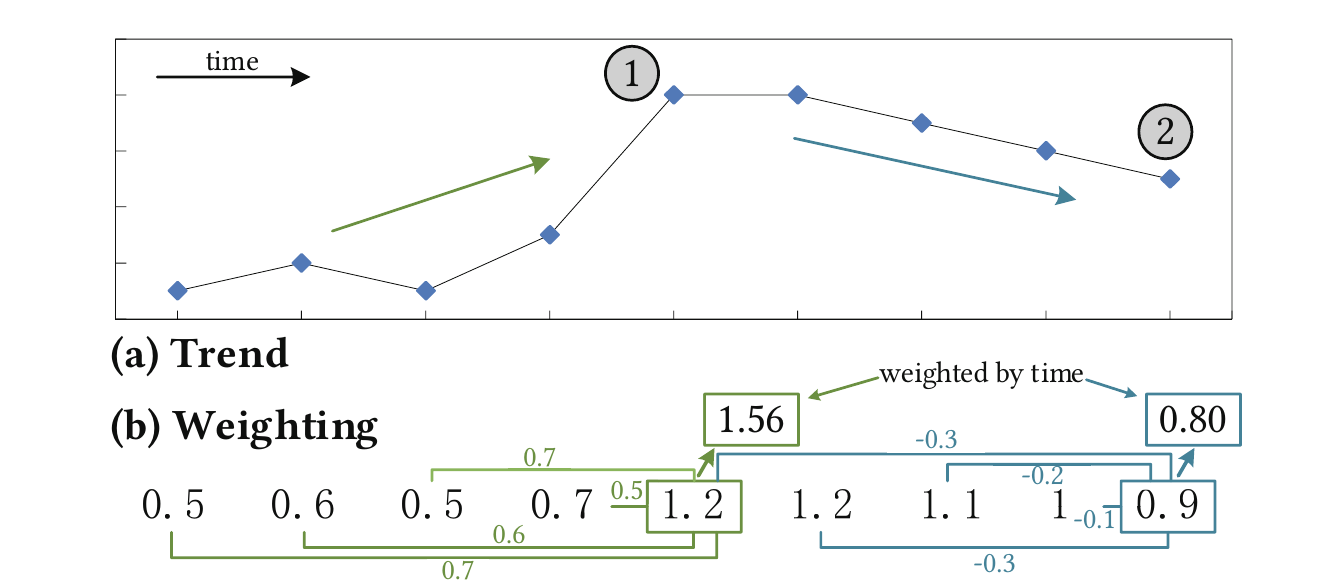}
			\caption{An example of weighted approach.}
			\label{fig_weightedApproach}
		\end{figure}

		\textbf{Weighted approach} uses temporal information to weight the regional divergence. Before adopting temporal information, each point at time slot $t$ should be allocated a divergence instead of using the regional divergence because the locations belong to a different region with the time flow. An approach is to assign the divergence of the region to each location contained by this region.
		\begin{align}
			\alpha_j^t = \mathfrak{D}_{r_{t,i}}, \quad j \in r_{t,i}, \label{eq_assign}
		\end{align}
		When each point at time slot $t$ owns a divergence, we consider the sequence of each point during a period $\left[t - \tau + 1, t\right]$. Figure \ref{fig_weightedApproach} illustrates the weighted approach. The fact of the divergence of a location goes up with time means the deviation of the point changes bigger and bigger. So we need to pay more attention to such a location. Consequently, such location should have a big weight, e.g., the point marked 1, which has weight 1.56. On the contrary, the decreasing trend of the divergence means the deviation changes smaller and smaller, e.g., the point marked 2, which has weight 0.80. It is reasonable to assign a small weight. Assuming the sequence of a point is $\alpha_i^{\left[t - \tau + 1, t\right]}$, we calculate the weight at time slot $t$ by:
		\begin{align}
			w_i^t &= \frac{2}{1+\exp(-\hat{\alpha}_i^{t})}, \\
			\hat{\alpha}_i^{t} &= \frac{1}{\tau}\sum\limits_{j=1}^{\tau-1}\vartheta^j \cdot (\alpha_i^t - \alpha_i^{t-j}) \label{eq_alpha}
		\end{align}
		where, $\vartheta \in (0, 1]$ is a decay factor. The value of $w_i^t$ has the following implication: $w_i^t>1$ means the deviation of the value is increscent, $w_i^t<1$ denotes the deviation of value is decrescent, and $w_i^t=1$ means the value is identical.

		After generating the weight of each point, we update the divergence of each region with:
		\begin{align}\label{eq_calculate_new_divergence}
			\mathfrak{D}_{r_{t,i}} = \frac{1}{\left| r_{t,i} \right|}\sum\limits_{j \in r_{t,i}}w_j^t \cdot \alpha_j,
		\end{align}

		For more robust detection results, an adaptively momentum-based method is used to generate the final abnormal regions:
		\begin{align}
			&mean^t =\jmath \cdot mean^{t-1} + (1-\jmath) \cdot mean\left( \mathfrak{D}_{r_{t}} \right), \label{eq_mean} \\
			&std^t = \jmath \cdot std^{t-1} + (1-\jmath) \cdot std\left( \mathfrak{D}_{r_{t}} \right), \label{eq_std} \\
			&\mathcal{A}_t = \left\{ r_{t,i} \left. \Big| \right. \mathfrak{D}_{r_{t,i}} \geq mean^t + \frac{mean^t}{r_{t,i}} \cdot std^t \right\}, \label{eq_anomaly_a}
		\end{align}
		where, $\jmath$ is a trade off between historical value and current value. $\mathfrak{D}_{r_{t}}$ means the divergence of all regions at time slot $t$. $mean(\cdot)$ and $std(\cdot)$ are the mean value function and the standard deviation function, respectively. Eq. (\ref{eq_anomaly_a}) is also used in \cite{Shi2017}.
		
		\textbf{Wavy approach} considers the fluctuation of the divergence series directly. Because the generated regions are dynamic at different time slot, we copy the regional divergence to each point similar to the weighted approach by Eq. (\ref{eq_assign}). The difference between the wavy approach and the weighted approach is the role of the temporal information. In the wavy approach, we directly detect anomalies on the divergence series instead of weighting.
		\begin{figure}[thbp]
			\centering
			\includegraphics[width=.78\textwidth]{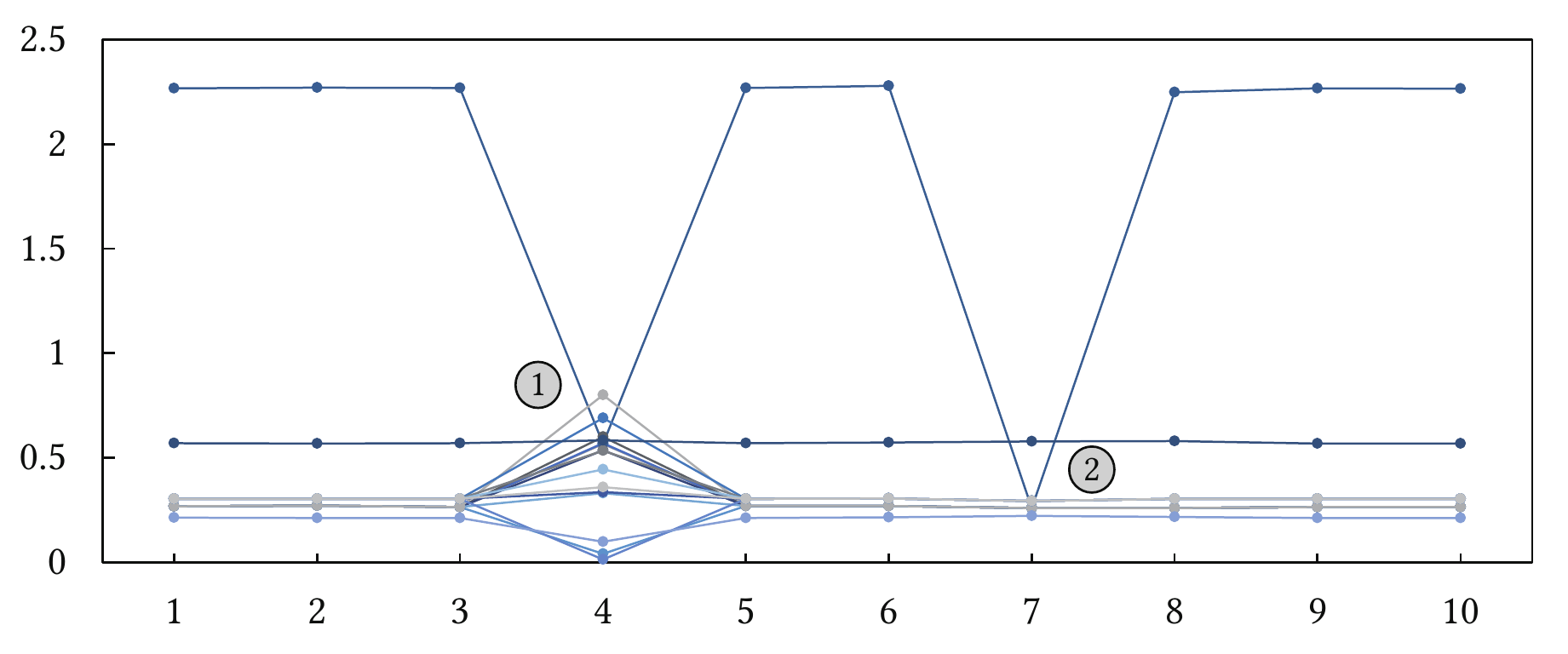} 
			\caption{Divergence series in wavy approach.}
			\label{fig_wavyApproach}
		\end{figure}

		Figure \ref{fig_wavyApproach} is an example of divergence series. Usually, the divergence of each point is stable. If there is a substantial fluctuation, we believe it is abnormal, e.g., the positions 1 and 2 of the top line. Though some existing methods can finish such a task, e.g., OC-SVM \cite{Schoelkopf1999} and skyline detection algorithm \cite{Boerzsoenyi2001}, we investigate a more simple and effective way to detect such fluctuation:
		\begin{align}
			&std_i^t = std \left( \alpha_i^{[t-\tau+1, t]} \right), \label{eq_std_b} \\
			&std^t = \jmath \cdot std^{t-1} + (1-\jmath) \cdot mean\left( STD^t \right), \\
			&mean_i^t =\jmath \cdot mean_i^{t-1} + (1-\jmath) \cdot mean\left( \alpha_i^{[t-\tau+1, t]} \right), \\
			&\mathcal{A}_{\alpha}^t = \left\{ \alpha_i^t \left. \Big| \right. \alpha_i^t \geq mean_i^t + \frac{mean_i^t}{\alpha_i^t} \cdot std^t \right\}, \label{eq_anomaly_b}
		\end{align}
		where, $STD^t = \left\{ std_1^t, std_2^t, \cdots, std_i^t, \cdots \right\}$ means the collection of current standard deviation from each point. $std^t$ is a global standard deviation. The $\jmath$, $mean(\cdot)$, and $std(\cdot)$ have the same meaning with Eq. (\ref{eq_mean}) and Eq. (\ref{eq_std}). After generating the point anomalies, we can aggregate the anomalous points to generate the anomalies $\mathcal{A}_t$ by the edges of Delaunay triangulation at time slot $t$.

		Algorithm \ref{alg_framework} summarizes the ReAD.
		\begin{algorithm}
			\caption{ReAD}\label{alg_framework}
			\KwIn{Dataset $s$, current time slot $t$, $d_c$ and $n_t(v)$ in CFDP, the number of location cluster $n_t(l)$, $\sigma$ in Gaussian kernel, detection approach $ap$, other paramenters $\lambda$, $\tau$, $\vartheta$, and $\jmath$}
			\KwOut{Anomalous regions $\mathcal{A}_t$}
			Construct the Delaunay triangulation $\mathcal{T}_{s_t}$ on $s_t$\;
			Clustering locations to $\mathcal{L}$ by (\ref{eq_rho}-\ref{eq_delta})\;
			Clustering readings to $\mathcal{V}$ by CFDP\;
			Partition $s_t$ to $r_t$ though $\mathcal{L} \cap \mathcal{V}$\;
			Calculate the divergence $\mathfrak{D}_{r_{t,i}}$ of each region using (\ref{eq_calculate_divergence})\;
			\eIf{$ap$ is weighted approach}
			{
				Generate the weights by (\ref{eq_assign}-\ref{eq_calculate_new_divergence})\;
				Generate anomalous regions $\mathcal{A}_t$ by (\ref{eq_mean}-\ref{eq_anomaly_a})\;
			}
			{
				Generate point anomalies $\mathcal{A}_\alpha^t$ by (\ref{eq_std_b}-\ref{eq_anomaly_b})\;
				Aggregate $\mathcal{A}_\alpha^t$ to $\mathcal{A}_t$ using the edges of $\mathcal{T}_{s_t}$\;
			} 
			\Return{$\mathcal{A}_t$}
		\end{algorithm}

		% ================>Experimental Setup<=====================
		\section{Experiments}
		\label{sec_experiments}
		In this section, we first evaluate the proposed ReAD on a synthetic dataset and compare it with state-of-the-art anomaly detection algorithms. Then, we perform two case studies on real-world datasets to demonstrate the feasibility of our approach in real-world applications.

		\subsection{Experiments Based on Synthetic Dataset}
		\label{sec_artificial_dataset}
		Due to there is no widely used standard benchmark for evaluating regional anomaly detection algorithms, we construct a synthetic dataset for quantitative analysis.
		\begin{figure}[htbp]
			\centering
			\subfloat[Synthetic spatial coordinates.] {
				\includegraphics[width=2.3in]{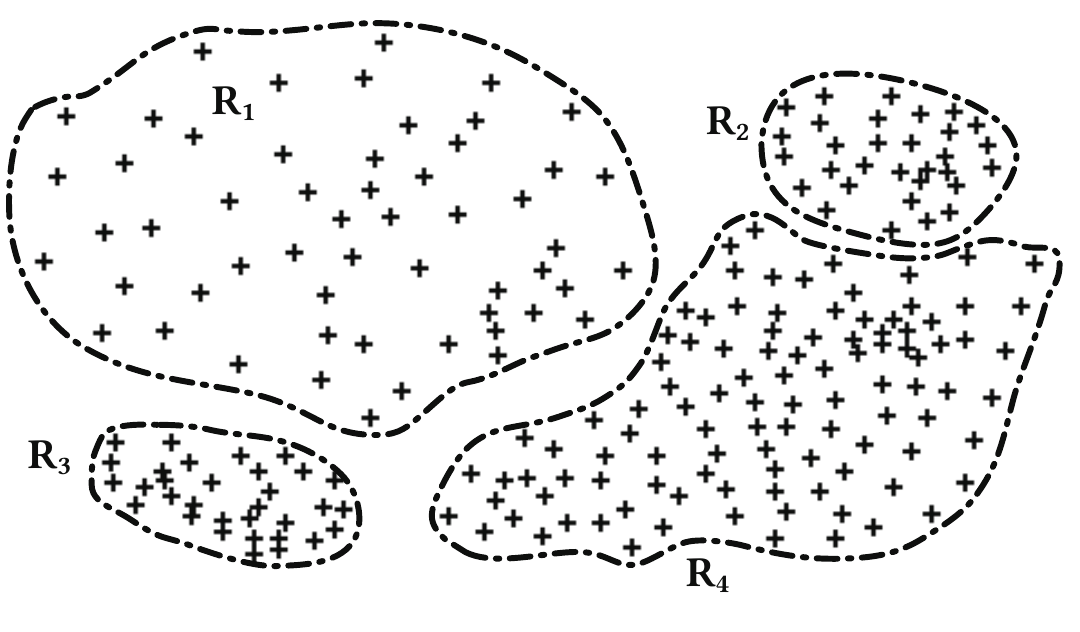}
				\label{fig_syntheticSpatioLocations}
			}
			\hspace{.1in}
			\subfloat[Base time-series.] {
				\includegraphics[width=3.3in]{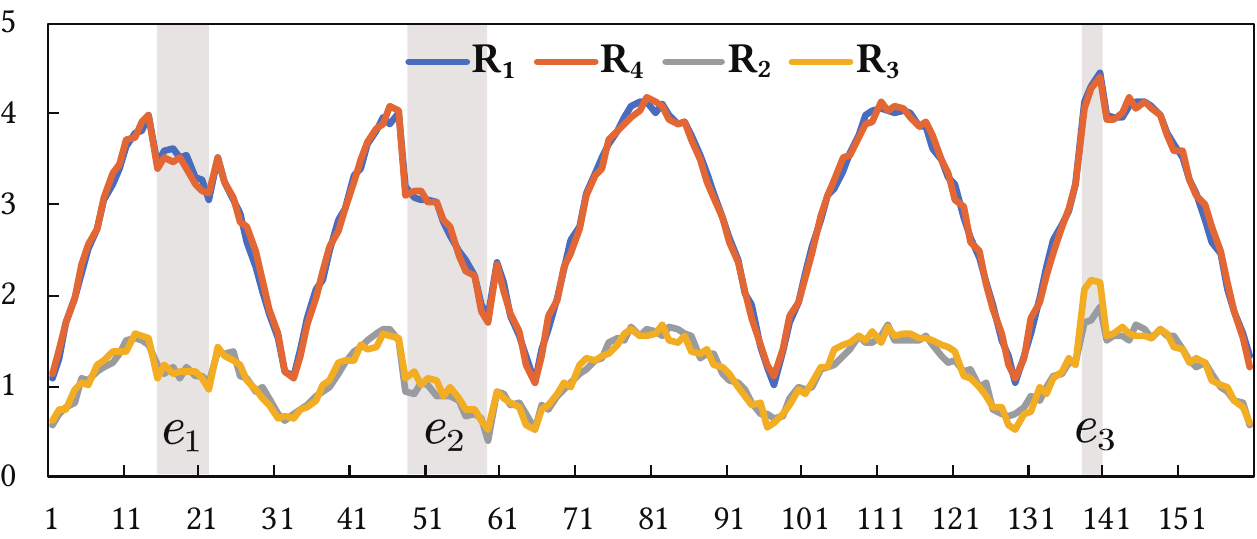}
				\label{fig_baseSeries}
			}
			\caption{Synthetic data generation.}
			\label{fig_syntheticData}
		\end{figure}

\begin{figure}[htbp]
			\centering
			\subfloat[An abnormal region.] {
				\includegraphics[width=2.3in]{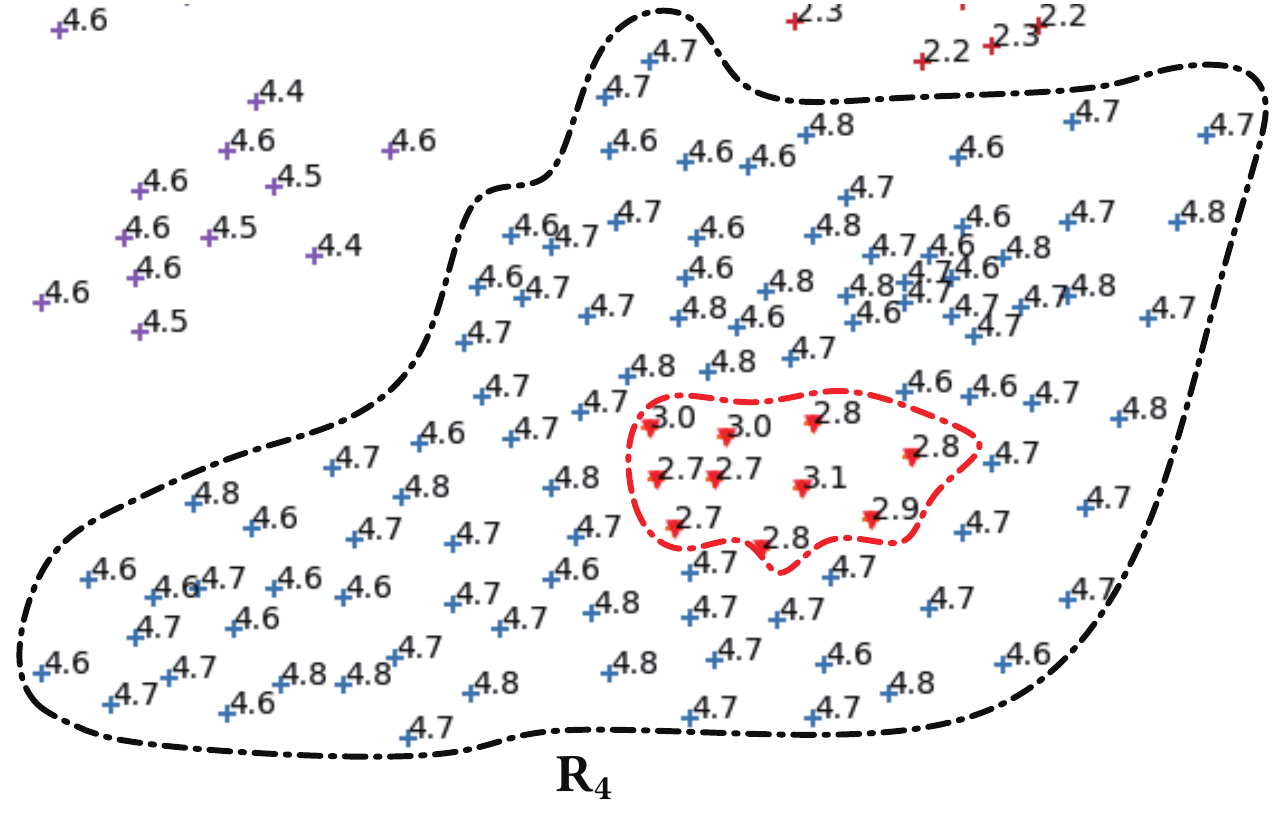}
				\label{fig_syntheticAnomalyLocations}
			}
			\hspace{.1in}
			\subfloat[An abnormal time-series.] {
				\includegraphics[width=2.4in]{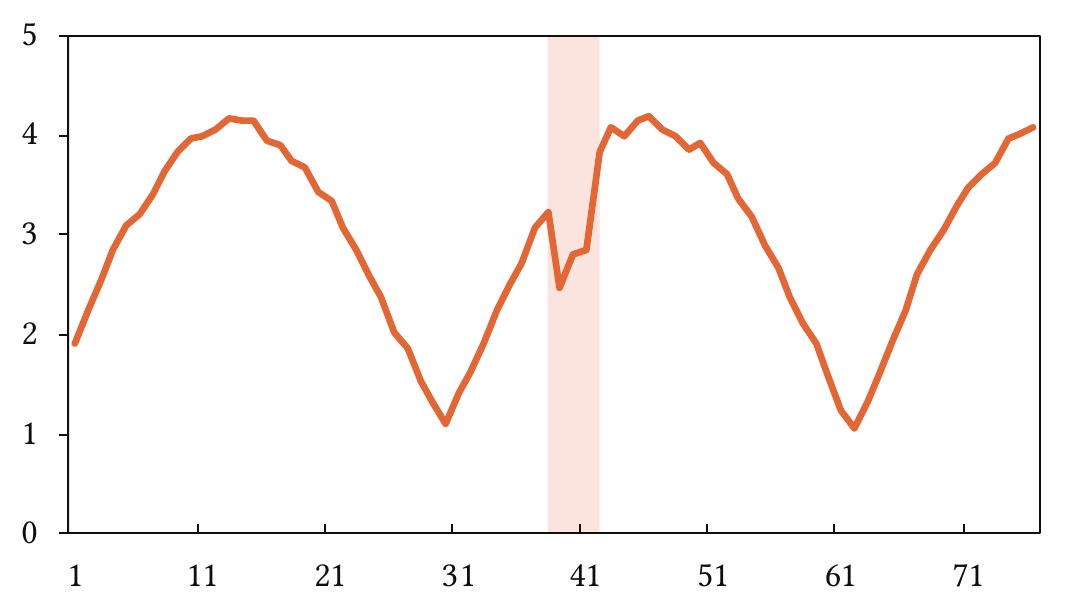}
				\label{fig_syntheticAnomalyTimeseries}
			}
			\caption{An example of injected anomaly.}
			\label{fig_syntheticAnomalyData}
		\end{figure}

		\textbf{Dataset Generation}. There are five steps to generate data source. \textbf{1) Generate four regions.} To mimic real spatial locations, we manually generate the latitude and longitude of each point by a GeoJSON generator\footnote{\url{http://geojson.io}}. As shown in Figure \ref{fig_syntheticSpatioLocations}, the generated spatial distribution is designed to present different densities, e.g., $R_1$ is sparser than other regions. Besides, each region has a different size and shape. \textbf{2) Generate time-series readings for each location.} The time-series is sampled from a base curve: $y_t=y_0 \cdot \left| \sin(x_t) \right| + b + g$, where $y_0$ is an amplitude, $b$ is a bias, and $g$ means a Gaussian noise. Figure \ref{fig_baseSeries} depicts two groups of base time-series we set for the synthetic coordinates. The upper group with $y_0=3$, $b=1$ is assigned to $R_1$ and $R_4$, and the lower group with $y_0=1$, $b=0.5$ belongs to $R_2$ and $R_3$. \textbf{3) Inject external influences.} We randomly inject some external influences to the base time-series, e.g., $e_1$, $e_2$, and $e_3$ in Figure \ref{fig_baseSeries}, by changing the base readings in the same trend (increase or decrease simultaneously). The external influences are set to mimic the influences caused by external factors, e.g., weather and holiday. \textbf{4) Generate readings by sampling.} We generate the time-series of each location from the base time-series of its belonging region by adding another Gaussian noise for each time. The Gaussian noise is set to mean 0 and standard deviation 0.5, and the sampling step is set to 0.1 in our experiments. Such generation type on time-series is a simplification from the actual scenario: typical periodicity and changing trend influenced by external factors, which also can be regarded as a microform of NYC bike dataset. \textbf{5) Inject anomalies.} After generating the time-series for each location, we randomly choice adjacent locations in the same region and a period to inject a defined anomaly. Specifically, we add to or subtract from the chosen series with a random, but constant value $\nu \in \left[1.5, 2.0\right]$. An example of injecting anomaly is shown as Figure \ref{fig_syntheticAnomalyData}. Summarily, the generated four regions contain 223 coordinates, and each of them involves a time-series with 6402 time-slots. Besides, the number of external influences is 82, and the number of injected anomalies is 200.

		\textbf{Baselines}. We compare ReAD with the following baselines, which are classical and widely used anomaly detection methods.
		\begin{itemize}
			\item \textbf{Hotelling's $T^2$}. It is a generalization of Student's t-statistic that is used in multivariate hypothesis testing \cite{MacGregor1995}.
			\item \textbf{RKDE}. It is the Robust Kernel Density Estimation \cite{Kim2012}. A Gaussian kernel with a standard deviation of 1.0 and the Hampel loss function are used.
			\item \textbf{LRT}. It fits a Poisson distribution on the historical data and uses likelihood ratio test as the anomaly score.
			\item \textbf{SAPRD} It designs multi-constrained graphs and local density to detect spatial anomaly regions \cite{Shi2017}.
			\item \textbf{MDI-KDE}. MDI is a Maximally Divergent Intervals algorithm \cite{Barz2018}.
		\end{itemize}
		
		Besides the baselines, we also show the effectiveness of both weighted approach and wavy approach of the ReAD. We use F1-Score with an Intersection over Union (IoU) as the metric to quantify the performance. It is a hit if the IoU has an overlap over 50\%.
		
		For Hotelling's $T^2$, RKDE, and LRT, we obtain regional detections from those point-wise baselines by gathering contiguous detections at each time slot based on the edges of the Delaunay triangulation constructed on spatial coordinates. For RKDE, we adopt a Gaussian kernel with a standard deviation of 1.0 and the Hampel loss function as the setting in \cite{Barz2018}. The LRT is set as \cite{Zheng2015}. The SAPRD and MDI-KDE keep the same parameters as they are used in \cite{Shi2017} and \cite{Barz2018}, respectively. The weighted approach and wavy approach are set with the same parameters to compare with other baselines. The numbers of location cluster $n_t(t)$ and value cluster $n_t(v)$ are set as 4 and 2, respectively. Besides, $\lambda$, $\vartheta$, and $\jmath$ are set to 1, and $\tau$ is set to 10. $\sigma$ is set by Scott's rule. Finally, the average of 10 runs with the same hyperparameters is reported.
		\begin{table}[htbp]
			\caption{\label{table_comparison} Performance comparison between baselines and our method on synthetic data.}
			\begin{center}
				\begin{tabular}{|c|c|c|c|}
					\hline
					\textbf{Methods} & \textbf{Precision} & \textbf{Recall} & \textbf{F1 Score} \\ 
					\hline
					Hotelling's $T^2$ & 0.16 & \textbf{0.96} & 0.27           \\ 
					RKDE & 0.06 & 0.72 & 0.10            \\
					LRT & 0.09 & 0.56 & 0.16            \\
					SAPRD & 0.10 & 0.60 & 0.17            \\
					MDI-KDE & 0.36 & 0.56 & 0.44            \\
					\hline
					\textbf{ReAD} (weighted) & 0.50 & 0.60 & 0.55            \\
					\textbf{ReAD} (wavy) & \textbf{0.71} & 0.68 & \textbf{0.69}            \\
					\hline
				\end{tabular}
			\end{center}
		\end{table}

		\textbf{Performance Analysis}. Table \ref{table_comparison} summarizes the detection results of different baselines and the proposed approaches. ReAD (weighted) denotes the weighted approach, and ReAD (wavy) denotes the wavy approach of the ReAD. They achieve F1-score gains over the baselines as the table shows. These improvements demonstrate the superior performance of the proposed ReAD methods. Specifically, the first three baselines perform poorly. The reason is they are point anomaly detection methods, which only utilize the temporal information and can not make good use of the spatial information. Likewise, SAPRD does not obtain a good score for the reason of losing temporal information. Although MDI-KDE considers spatial and temporal information simultaneously, it is still worse than ReAD because its spatial partition is always rectangular. On the contrary, the proposed ReAD partition the regions with arbitrary shapes. Compared with the ReAD (weighted), ReAD (wavy) has a better performance. It proves spatio-temporal anomaly detection is more effective than spatial anomaly detection in this synthetic task. We also notice that Hotelling’s $T^2$ and RKDE have very high recall values but low precision values. The reason is they radically regard most of the candidates as anomalies. Otherwise, they will get close 0 precision values. However, our methods have a relative balance on precision and recall, thus achieve better F1 scores.
		\begin{figure}[thbp]
			\centering
			\includegraphics[width=.55\textwidth]{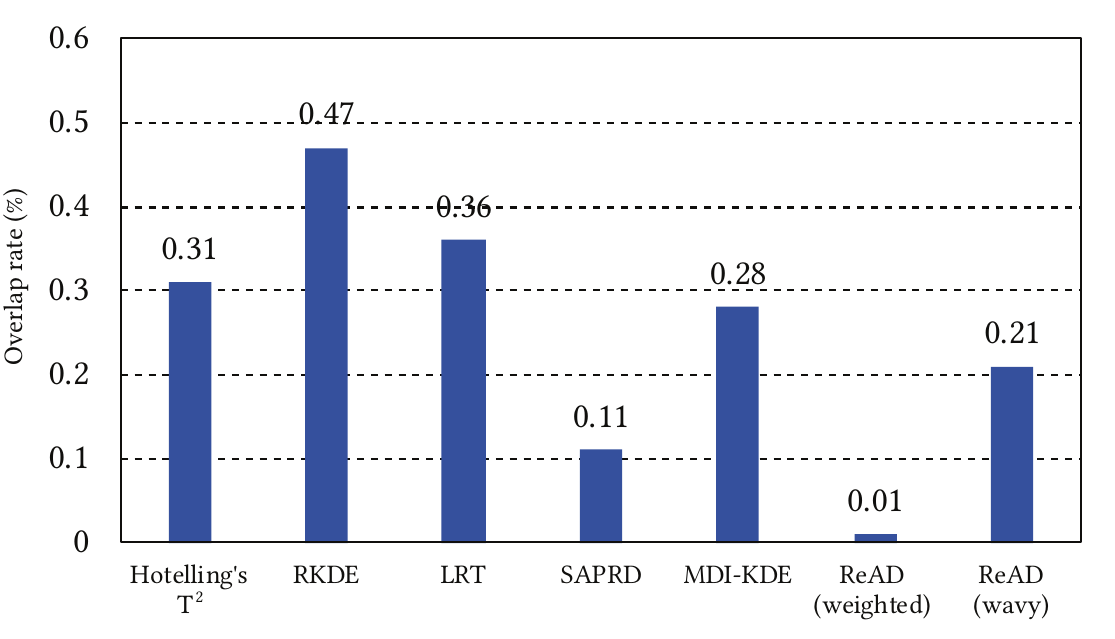}
			\caption{Robustness comparison between baselines and our method on synthetic data. The value means the ratio of the detected anomalies which overlap with the external influence in whole detected anomalies.}
			\label{fig_synheticAnomalyDetectedRatioOfExternal}
		\end{figure}
		
		\textbf{Robustness Analysis}. We further design an experiment to compare the robustness of our framework. The ratio of detected anomalies overlapping with the external influence is compared — the lower of the ratio, the better of the robustness. As can be seen in Figure \ref{fig_synheticAnomalyDetectedRatioOfExternal}, ReAD is capable of finding anomalies even with external influences. ReAD (weighted) and SAPRD have a lower ratio than others because they do not care about the continuity of time but detect spatial anomalies. However, others focus on temporal information, so they are deeper influenced by external changes. Nevertheless, ReAD (weighted) and ReAD (wavy) have better performance than others, which benefits from the dynamic region partition and relative divergence.

		\textbf{Discussion}. Although we have generated data source manually and designed a quantitative experiment to evidence the effectiveness of the proposed ReAD, it is not very convincing that the ReAD can always outperform other anomaly detection methods. One reason is the data synthesis process may implicitly adopt the same assumption of the proposed method. Another reason is we can not perform a real quantitative evaluation on a real-life dataset due to a lack of labels. This situation can happen in almost any unsupervised detection framework \cite{Zheng2015,Zhang2018}. It also reflects the fact that the definition and solution of anomaly detection vary from scenario to scenario. Nevertheless, we can argue that the proposed ReAD is still a solution for the regional anomaly detection and work well under a certain situation. Beside, building a real-life quantitative evaluation will be as our future work. Instead of quantitative evaluations on real-life datasets, the following content shows two case studies to demonstrate the feasibility of our approach.

		\subsection{Case Study I: Credit Detection of Companies}
		\label{sec_case_i}
		In the case, the weighted approach of ReAD is applied to detect the credit anomaly of companies. The dataset collected from a city of China contains company names and their credit scores. Typically, the credit score of a company does not change or change with a small value within a long period. Thus, there is nothing wrong with a company when checking the fluctuation of the score-series if it always keeps a lower score. On the contrary, considering the credit score of other companies around it is more useful. That is what the weighted approach wants to address. The dataset contains 3434 companies. The parameters are set to $n_t(l)=6$, $n_t(v)=3$, $\lambda=\tau=1$, $\vartheta=\jmath=0$ because just one time slot is involved, and $\sigma$ is set by the Scott's rule.
		\begin{figure}[thbp]
			\centering
			\includegraphics[width=.68\textwidth]{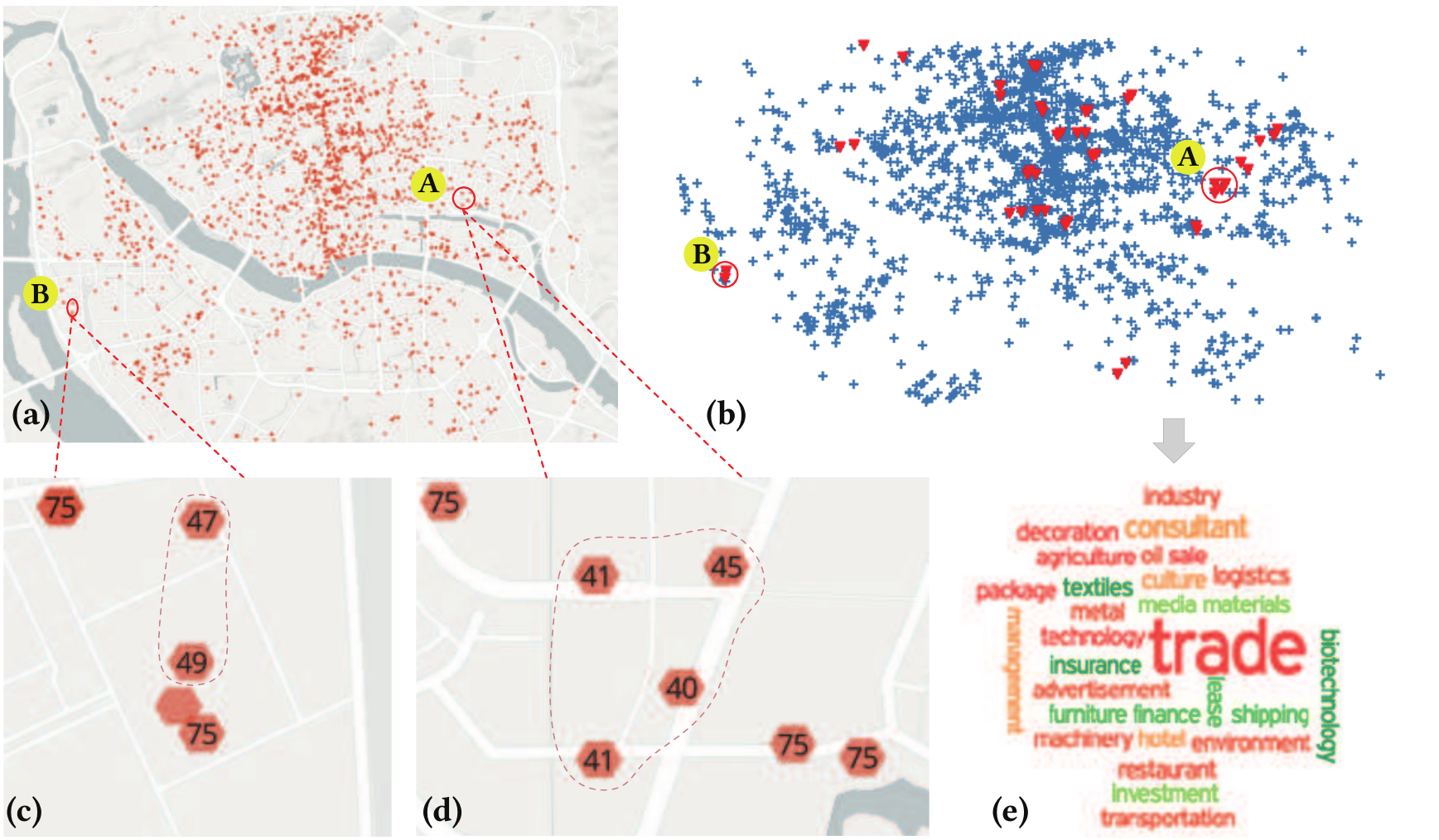}
			\caption{The potential anomaly pattern about a city of China.}
			\label{fig_caseCreditResults}
		\end{figure}

		Figure \ref{fig_caseCreditResults} presents the detection results. Figure \ref{fig_caseCreditResults}b is the detection results with top 20 anomalous regions marked with red triangles. As expected, regions with quite low credit scores are recognized by the proposed ReAD method, e.g., A and B in Figure \ref{fig_caseCreditResults}c and Figure \ref{fig_caseCreditResults}d, respectively. A word cloud is generated with the names of the detected companies. The visualization is shown as Figure \ref{fig_caseCreditResults}e, in which the ``trade''  indicates the trade industry has a potential poor credit. The case proves the capability of our method to do decision making and supervision.

		Our framework is designed based on the regional partition, which is meaningful in the above scenario. Because each region has different economic conditions, the companies in different regions have different credit scores. Thus, the collected score is heterogeneous on spatial distribution. The regional detection could find the clusters of low-credit companies, and further use to diagnosis industry credit and manage regional economic.

		\subsection{Case Study II: Events Detection in New York City}
		\label{sec_case_ii}
\begin{figure}[htbp]
			\centering
			\includegraphics[width=.78\textwidth]{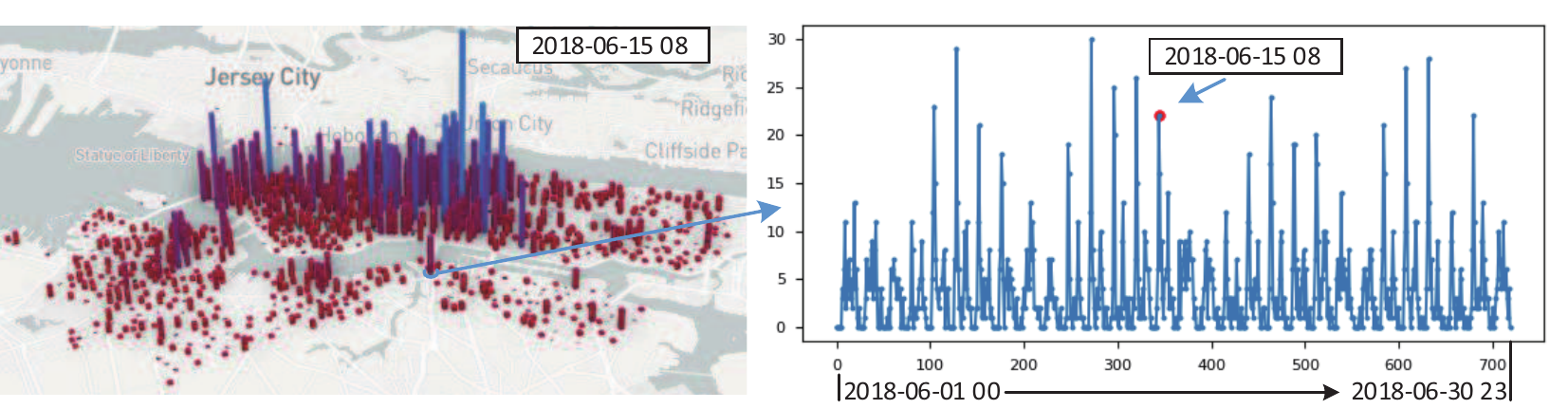}
			\caption{NYC bike dataset.}
			\label{fig_spatio_temporal_attribute}
		\end{figure}

		To demonstrate the feasibility of the wavy approach for real problem, we also perform events detection in New York City using its bike rental data, which is generated by the bike-sharing system\footnote{\url{https://www.citibikenyc.com/system-data}}. We count the hourly check-out of each station from 6/1/2018 to 8/31/2018, which has 808 bike stations and 2208 time slots. Figure \ref{fig_spatio_temporal_attribute} is an exmaple of distribution of the bike stations (the left part) and a check-out series of a staion from 6/1/2018 to 6/30/2018(the right part). We evaluate our method by the corresponding events reported by \textit{nycinsiderguide}\footnote{\url{https://www.nycinsiderguide.com/}} refering to the study of \cite{Zheng2015,Zhang2018}. The details of the 15 primarily reported events are listed in Table \ref{table_events}. The information in this table is collected manually.

		Table \ref{table_comparison_bike} presents the hit events of our method and different baselines. The top 15 anomalies detected by these models are checked. The `hit' means the detected results have an overlap with any address and corresponding start or end time in Table \ref{table_events}. The parameters are set with the same as the experiments designed on the synthetic dataset. ReAD (wavy) hits 6 events and other methods hit less than or equal with 4 events. It shows that our method has an advantage of detecting real-world events without knowing any other external influences, e.g., weather. It indicates that using regional partition and the divergence in ReAD (wavy) works well on spatio-temporal anomaly detection.
		\begin{table}[htbp]
			\caption{\label{table_comparison_bike} Events hit on NYC bike dataset.}
			\begin{center}
				\begin{tabular}{|p{3cm}|p{2.5cm}|cl}
					\hline
					\textbf{Methods} & \textbf{Hit Event IDs} \\ 
					\hline\hline
					Hotelling's $T^2$  &    8        \\ 
					RKDE &     None       \\ 
					LRT  &     12       \\ 
					SAPRD &    3, 12        \\ 
					MDI-KDE &  3, 6, 12, 15          \\ 
					\hline\hline
					\textbf{ReAD} (weighted) &   4, 6, 12, 13         \\ 
					\textbf{ReAD} (wavy) &  3, 4, 6, 8, 12, 13          \\ 
					\hline
				\end{tabular}
			\end{center}
		\end{table}
		\begin{table*}[htbp]
			\caption{\label{table_events} Events reported by \textit{nycinsiderguide.com}}
			\begin{center}
				\begin{tabular}{|p{0.2cm}|p{6.5cm}|p{3.2cm}|p{2cm}|p{2.1cm}|}
					\hline
					& \textbf{Event Name} & \textbf{Address} & \textbf{Start Time} & \textbf{End Time} \\ 
					\hline
					1 & Governors Ball Music Festival & Randall's Island & 06/01 11:45AM & 06/01 \\
					2 & Taste of Times Square & 46th Street & 06/04 5:00PM & 06/04 9:00PM \\
					3 & Central Park Taste of Summer & Central Park & 06/06 7:00PM & 06/06 11:00PM \\
					4 & AVP(Volleyball) New York City Open & Tribeca Courts & 06/07 8:00AM & 06/10 6:00PM \\
					5 & National Puerto Rican Day Parade & Fifth Avenue, 44th to 79th Street & 06/10 11:00AM & 06/10 5:00PM \\
					6 & Museum Mile Festival & Fifth Ave, 82nd to 105th Street & 06/12 6:00PM & 06/12 9:00PM \\
					7 & NY Philharmonic Concerts in the Parks & Central Park & 06/13 8:00PM & 06/13 9:30PM \\
					8 & American Crafts Festival at Lincoln Center & Lincoln Center & 06/14 10:00PM & 06/14 7:00PM \\
					9 & Bryant Park Film Festival & Bryant Park & 06/18 8:00PM & 06/18 9:00PM \\
					10 & Macy’s July 4th Fireworks & East River & 07/04 9:25PM & 07/04 \\
					11 & FREE Broadway in Bryant Park & Lawn 42nd St and 6th Ave & 07/06 12:30PM & 08/10 1:30PM \\
					12 & Riverflicks FREE Movies in Hudson River Park & Adults: Pier 63 at W. 23rd Street & 07/11 8:30PM & 07/11 \\
					13 & Riverflicks FREE Movies in Hudson River Park & Kids: Pier 46 at Charles Street & 07/13 8:30PM & 07/13 \\
					14 & Jay Z \& Beyoncé OTR II & MetLife Stadium & 08/05 7:30PM & 08/05 \\
					15 & Battery Dance Festival NYC & Battery Park City & 08/11 7:00PM & 08/17 9:00PM \\
					\hline
				\end{tabular}
			\end{center}
		\end{table*}

		% ================>Related work<=====================
		\section{Related Work}
		\label{sec_related_work}
		\subsection{Anomaly Detection}
		As a critical problem in practical applications, anomaly detection has been studied extensively in the past decades \cite{Chandola2009,Atluri2018}. In urban computing, the form of data is usually spatio-temporal data, in which the readings are influenced by many factors, e.g., location, time, and weather, etc. Thus, these anomaly detection methods for univariate time series \cite{Wei2005,Siffer2017}, multivariate time series \cite{Takeishi2014}, and spatial outliers \cite{Shi2016,Zheng2017} are unsuitable to the urban scenario. The main reason is they ignore the joint representation of spatial and temporal attributes. In this paper, we focus on urban anomaly detection.
		
		A direction to explore the urban anomaly detection is to use the movement of the observations (e.g., bike, taxi, and crowd). Because these movements generate a lot of trajectories every day, trajectory anomalies detection is a direct way in urban anomaly detection. \cite{Ge2010} computed the average direction of each spatial grid and identified an anomalous trajectory if its covered grids exist some unexpected behaviors. In addition to the above two approaches. \cite{Liu2011} proposed a graph-based method to glean the problematic design in urban planning. Besides, some supervised methods are also explored to detect anomalous trajectory shapes \cite{Li2006,Li2007}.
		
		Considering the information of the observations from the statistical perspective, another direction to investigate the urban anomaly detection is to exploit the collected readings of spatial locations on time interval. Different from the trajectory anomalies detection addressed by line-based methods, group anomalies (or collective anomalies, regional anomalies) are what point-based or region-based approach focuses. \cite{Zheng2015} first partition a city into some regions by major roads, such as highways and arterial roads. The generated regions are the minimal unit to extract statistic features, like taxi amount or bike amount at each time interval. Then, detected the anomalies on a probabilistic model. \cite{Zhang2018} used the same partition strategy as \cite{Zheng2015} and identified anomalies by the changes of similarity measurement (Pearson Correlation Coefficient) between the target region and other historically similar regions. We focus on regional anomalies detection in this paper. Different from previous approaches built on the fixed partition, the proposed framework is designed based on the dynamic region partition, which alleviates the sparse and heterogeneity of data distribution from the beginning.
		
		\subsection{Regional Partition}
		The dynamic region partition involves dynamic clusters. Despite many clustering algorithms including K-means \cite{MacQueen1967} and DBSCAN \cite{Ester1996} can group similar points to several clusters, they do not address the fundamental necessities of our regional partition: arbitrarily shape and full-coverage, and do not take into account spatial information exactly. \cite{Chen2015} first identified the set of core locations across all timestamps and then grew around the core locations at every time stamps to capture the dynamic behavior occurring at the boundaries. However, the clusters are also not full-coverage clusters. \cite{Shi2016} used a multilevel and constrained Delaunay triangulation to detect spatial point event outliers. Different from their cutting approach between nodes, we cluster locations through the definition of node density. Such density definition can avoid the error caused by the adaptive coefficient and reduce the cost of computation.

		% ================>Conclusion<=====================
		\section{Conclusion}
		\label{sec_conclusion}
		In this paper, we propose a novel framework ReAD to detect regional anomalies. The framework partitions regions according to coordinates proximity and readings similarity on spatial locations and non-spatial attributes at each time slot. To cluster the geographical locations, we propose a clustering algorithm based on Delaunay triangulation. Then, the abnormal metric of each region is calculated by the relative divergence. Based on the divergence, two types of regional anomaly detection approach are proposed to address different scenarios which use temporal information differently. We finally evaluate the proposed framework on the elaborate synthetic dataset and other two real-world applications. They all demonstrate the effectiveness and feasibility of the proposed framework.
		
		% ================>Bibliography<====================
		\bibliographystyle{unsrt}  
		\bibliography{RegionalAnomalyDetection-Ref}

\begin{thebibliography}{10}

\bibitem{Chandola2009}
Varun Chandola, Arindam Banerjee, and Vipin Kumar.
\newblock Anomaly detection: {A} survey.
\newblock {\em {ACM} Comput. Surv.}, 41(3):15:1--15:58, 2009.

\bibitem{Zheng2014}
Yu~Zheng, Licia Capra, Ouri Wolfson, and Hai Yang.
\newblock Urban computing: concepts, methodologies, and applications.
\newblock {\em ACM Transactions on Intelligent Systems and Technology (TIST)},
  5(3):38, 2014.

\bibitem{Zhang2018}
Huichu Zhang, Yu~Zheng, and Yong Yu.
\newblock Detecting urban anomalies using multiple spatio-temporal data
  sources.
\newblock {\em Proceedings of the ACM on Interactive, Mobile, Wearable and
  Ubiquitous Technologies}, 2(1):54, 2018.

\bibitem{Shi2016}
Yan Shi, Min Deng, Xuexi Yang, and Qiliang Liu.
\newblock Adaptive detection of spatial point event outliers using multilevel
  constrained delaunay triangulation.
\newblock {\em Computers, Environment and Urban Systems}, 59:164--183, 2016.

\bibitem{Zheng2017}
Guanjie Zheng, Susan~L. Brantley, Thomas Lauvaux, and Zhenhui Li.
\newblock Contextual spatial outlier detection with metric learning.
\newblock In {\em KDD}, pages 2161--2170. {ACM}, 2017.

\bibitem{Ge2010}
Yong Ge, Hui Xiong, Zhi{-}Hua Zhou, Hasan~Timucin Ozdemir, Jannite Yu, and
  Kuo~Chu Lee.
\newblock Top-eye: top-k evolving trajectory outlier detection.
\newblock In {\em CIKM}, pages 1733--1736. {ACM}, 2010.

\bibitem{Liu2011}
Wei Liu, Yu~Zheng, Sanjay Chawla, Jing Yuan, and Xing Xie.
\newblock Discovering spatio-temporal causal interactions in traffic data
  streams.
\newblock In {\em Proceedings of the 17th {ACM} {SIGKDD} International
  Conference on Knowledge Discovery and Data Mining}, pages 1010--1018. {ACM},
  2011.

\bibitem{Li2006}
Xiaolei Li, Jiawei Han, and Sangkyum Kim.
\newblock Motion-alert: Automatic anomaly detection in massive moving objects.
\newblock In {\em International Conference on Intelligence and Security
  Informatics}, volume 3975 of {\em Lecture Notes in Computer Science}, pages
  166--177. Springer, 2006.

\bibitem{Li2007}
Xiaolei Li, Jiawei Han, Sangkyum Kim, and Hector Gonzalez.
\newblock {ROAM:} rule- and motif-based anomaly detection in massive moving
  object data sets.
\newblock In {\em Proceedings of the Seventh {SIAM} International Conference on
  Data Mining}, pages 273--284. {SIAM}, 2007.

\bibitem{Zheng2015}
Yu~Zheng, Huichu Zhang, and Yong Yu.
\newblock Detecting collective anomalies from multiple spatio-temporal datasets
  across different domains.
\newblock In {\em Proceedings of the 23rd {SIGSPATIAL} International Conference
  on Advances in Geographic Information Systems}, pages 2:1--2:10. {ACM}, 2015.

\bibitem{Zhang2017}
Junbo Zhang, Yu~Zheng, and Dekang Qi.
\newblock Deep spatio-temporal residual networks for citywide crowd flows
  prediction.
\newblock In {\em AAAI}, pages 1655--1661, 2017.

\bibitem{Shi2017}
Yan Shi, Min Deng, Xuexi Yang, and Qiliang Liu.
\newblock A spatial anomaly points and regions detection method using
  multi-constrained graphs and local density.
\newblock {\em Transactions in GIS}, 21(2):376--405, 2017.

\bibitem{Barz2018}
Bjorn Barz, Erik Rodner, Yanira~Guanche Garcia, and Joachim Denzler.
\newblock Detecting regions of maximal divergence for spatio-temporal anomaly
  detection.
\newblock {\em IEEE Transactions on Pattern Analysis and Machine Intelligence},
  2018.

\bibitem{Ester1996}
Martin Ester, Hans-Peter Kriegel, J{\"o}rg Sander, and Xiaowei Xu.
\newblock A density-based algorithm for discovering clusters in large spatial
  databases with noise.
\newblock In {\em KDD}, volume~96, pages 226--231, 1996.

\bibitem{Rodriguez2014}
Alex Rodriguez and Alessandro Laio.
\newblock Clustering by fast search and find of density peaks.
\newblock {\em Science}, 344(6191):1492--1496, 2014.

\bibitem{Tsai1993}
Victor~JD Tsai.
\newblock Delaunay triangulations in tin creation: an overview and a
  linear-time algorithm.
\newblock {\em International Journal of Geographical Information Science},
  7(6):501--524, 1993.

\bibitem{Liu2013}
Song Liu, Makoto Yamada, Nigel Collier, and Masashi Sugiyama.
\newblock Change-point detection in time-series data by relative density-ratio
  estimation.
\newblock {\em Neural Networks}, 43:72--83, 2013.

\bibitem{Jiang2015}
Meng Jiang, Alex Beutel, Peng Cui, Bryan Hooi, Shiqiang Yang, and Christos
  Faloutsos.
\newblock A general suspiciousness metric for dense blocks in multimodal data.
\newblock In {\em 2015 {IEEE} International Conference on Data Mining}, pages
  781--786. {IEEE} Computer Society, 2015.

\bibitem{Schoelkopf1999}
Bernhard Sch{\"{o}}lkopf, Robert~C. Williamson, Alexander~J. Smola, John
  Shawe{-}Taylor, and John~C. Platt.
\newblock Support vector method for novelty detection.
\newblock In {\em NIPS}, pages 582--588. The {MIT} Press, 1999.

\bibitem{Boerzsoenyi2001}
Stephan B{\"{o}}rzs{\"{o}}nyi, Donald Kossmann, and Konrad Stocker.
\newblock The skyline operator.
\newblock In {\em ICDE}, pages 421--430. {IEEE} Computer Society, 2001.

\bibitem{MacGregor1995}
John~F MacGregor and Theodora Kourti.
\newblock Statistical process control of multivariate processes.
\newblock {\em Control Engineering Practice}, 3(3):403--414, 1995.

\bibitem{Kim2012}
JooSeuk Kim and Clayton~D Scott.
\newblock Robust kernel density estimation.
\newblock {\em Journal of Machine Learning Research}, 13(Sep):2529--2565, 2012.

\bibitem{Atluri2018}
Gowtham Atluri, Anuj Karpatne, and Vipin Kumar.
\newblock Spatio-temporal data mining: {A} survey of problems and methods.
\newblock {\em {ACM} Comput. Surv.}, 51(4):83:1--83:41, 2018.

\bibitem{Wei2005}
Li~Wei, Nitin Kumar, Venkata~Nishanth Lolla, Eamonn~J. Keogh, Stefano Lonardi,
  and Chotirat~(Ann) Ratanamahatana.
\newblock Assumption-free anomaly detection in time series.
\newblock In {\em SSDBM}, pages 237--240, 2005.

\bibitem{Siffer2017}
Alban Siffer, Pierre{-}Alain Fouque, Alexandre Termier, and Christine
  Largou{\"{e}}t.
\newblock Anomaly detection in streams with extreme value theory.
\newblock In {\em KDD}, pages 1067--1075. {ACM}, 2017.

\bibitem{Takeishi2014}
Naoya Takeishi and Takehisa Yairi.
\newblock Anomaly detection from multivariate time-series with sparse
  representation.
\newblock In {\em {IEEE} International Conference on Systems, Man, and
  Cybernetics}, pages 2651--2656. {IEEE}, 2014.

\bibitem{MacQueen1967}
James MacQueen et~al.
\newblock Some methods for classification and analysis of multivariate
  observations.
\newblock In {\em Proceedings of the fifth Berkeley symposium on mathematical
  statistics and probability}, volume~1, pages 281--297. Oakland, CA, USA,
  1967.

\bibitem{Chen2015}
Xi~C Chen, James~H Faghmous, Ankush Khandelwal, and Vipin Kumar.
\newblock Clustering dynamic spatio-temporal patterns in the presence of noise
  and missing data.
\newblock In {\em IJCAI}, pages 2575--2581, 2015.

\end{thebibliography}
		
	\end{document}